\documentclass{article}
\usepackage[dvips]{graphicx}
\title{Method of determination of the most probable coordinate of formation of $\alpha$-particle in $\alpha$-decay}

\author{Sergei~P.~Maydanyuk\thanks{E-mail: maidan@kinr.kiev.ua},
Sergei~V.~Belchikov\thanks{E-mail: sbelchik@kinr.kiev.ua}\\
\small\emph{Institute for Nuclear Research, National Academy of Sciences of Ukraine,} \\
\small\emph{prosp. Nauki, 47, Kiev-28, 03680, Ukraine}}

\date{\small\today}

\begin{document}

\maketitle

\begin{abstract}
Method of multiple internal reflections (method MIR) in description of $\alpha$-decay of nucleus in the spherically symmetric approximation is presented in paper. In approach MIR the formalism of calculation of amplitudes of wave function, described moving of $\alpha$-particle from the internal region outside with its tunneling through a realistic radial $\alpha$-nucleus barrier of arbitrary shape, has been constructed at first time.
The method MIR gives convergent values for the amplitudes of transmission and reflection, coefficients of penetrability and reflection, obtained in description of leaving of the $\alpha$-particle relatively a potential with barrier in form of a number of rectangular steps (multi-steps potential), with tending this potential to the realistic $\alpha$-nucleus one, and limit values of the amplitudes and coefficients are considered as exact concerning the realistic $\alpha$-nucleus potential (without application of WKB approximation).
In approach MIR a dependence of the penetrability coefficient and half-live on the location of the starting point, from where $\alpha$-particle begins to move outside, is opened. From here, we define a radial coordinate of the most probable formation of the $\alpha$-particle inside nucleus before its $\alpha$-decay as such starting point, at moving of the $\alpha$-particle from which outside the half-live, calculated by approach MIR, is maximally closed to its experimental value.
For isotopes ${\rm Po}$ with different mass number $A$ we have obtained essentially closer values for half-live, calculated by such approach, to their experimental values in a comparison with half-lives obtained by approach WKB.
This result is stable for all studied nuclei.
\end{abstract}





\section{Introduction
\label{sec.intro}}

An approach for description of one-dimensional motion of a non-relativistic particle above a barrier on the basis of consideration of multiple internal reflections of stationary waves relatively boundaries of this barrier, described a motion of this particle inside the barrier region, was considered in a number of papers and is known \cite{Fermor.1966.AJPIA,McVoy.1967.RMPHA,Anderson.1989.AJPIA}.
In such approach it needs in expression of wave function (WF) in the barrier region to separate components, having fluxes directed to opposite sides. For a problem of the motion of the particle above a rectangular barrier these components are found naturally --- as plane waves $e^{\pm ikx}$ where $k$ is wave vector. To obtain solutions for description of tunneling of this particle under such barrier it turns out more difficult, because in consideration of tunneling as stationary process the fluxes defined on the basis of decreasing and increasing parts of the stationary WF in dependence on $x$ inside the barrier region (which are analytical continuations of the corresponding waves for case of above-barrier propagation) equal to zero and it is not corrected to consider such ``waves'' as propagated (or tunneled) under the barrier from physical point of view. If to define the waves under the barrier by another way (for example, at first to require existence of non-zero fluxes in the barrier region in each step and then to find solutions of the waves), then we cannot obtain continuity in transition of expressions for the waves for the above-barrier and sub-barrier cases, and total amplitudes will be different on the corresponding amplitudes, determined by standard approach of quantum mechanics as, for example, in~\cite{Landau.v3.1989} (see p.~102--106; for simplicity we shall further name such approach for determination of the amplitudes of the stationary wave function as \emph{standard (direct) method of quantum mechanics}). However, the flux, defined on the basis of the total stationary WF, is non-zero, that points out that tunneling of the particle under the barrier exists. So, there was a difficulty in correct description of tunneling of the particle under the barrier on the basis of multiple internal reflections of the stationary waves.

However, this question was successfully resolved at first time in papers~\cite{Maydanyuk.2000.UPJ,Maydanyuk.2002.JPS} (see also~\cite{Maydanyuk.2003.PhD-thesis}, part~1), in result of introduction of wave packets (WPs). In such approach, tunneling of the particle under the barrier is considered on the basis of multiple internal reflections of the wave packets relatively boundaries of the barrier (this approach was named as \emph{method of multiple internal reflections} or \emph{method MIR}). In such approach it succeeded in connecting:
\begin{itemize}

\item
continuous transition of solutions for separate WPs concerning each reflection from multiple ones (i.~e. in each step, according to formalism of method MIR), total WPs (and corresponding stationary WFs) in transition from the above-barrier motion to the under-barrier tunneling and back;

\item
coincidence of the transmitted and reflected total amplitudes of stationary part of WF (in each spacial region of the potential) in approach MIR with the corresponding amplitudes obtained by standard method of quantum mechanics;

\item
all fluxes, constructed on the basis of WPs in each step, are non-zero, that allows to consider propagation of multiple packets under the barrier (i.~e. their ``tunneling'').
\end{itemize}
As a further development of a non-stationary formalism for description of tunneling processes in direction of papers~\cite{Olkhovsky.1992.PRPLC,Olkhovsky.1997.Trieste}, this method has own description (interpretation) of non-stationary tunneling of the particle, allowing to study this process at interesting time moment or space point. In calculation of phase times this method has shown as enough simple and convenient \cite{Maydanyuk.2006.FPL}. The formalism of MIR was developed also for description of scattering of the particle on nucleus and $\alpha$-decay in the spherically symmetric approximation with the simplest radial barriers in~\cite{Maydanyuk.2000.UPJ,Maydanyuk.2002.JPS} (see also~\cite{Maydanyuk.2003.PhD-thesis}, part~2), and for tunneling of photons (for example, see~\cite{Maydanyuk.2002.JPS,Maydanyuk.2006.FPL}).

However, in papers~\cite{Fermor.1966.AJPIA,McVoy.1967.RMPHA,Anderson.1989.AJPIA,Maydanyuk.2002.JPS,Maydanyuk.2006.FPL} correctness of applicability of the multiple reflections was proved for description of the motion of the particle above the barrier on the basis of waves, and also for description of tunneling of the particle under the barrier on the basis of wave packets, if this barrier has a shape of one rectangular step. But, addition of the second step to such barrier complicates essentially consideration of the multiple reflections of packets (or waves) relatively all boundaries and search of exact solutions for the amplitudes. It becomes unclear how to separate the needed internal reflections from all their variety inside interesting spacial part of the barrier for determination of the amplitudes here (for example, if the barrier consists of arbitrary number of the rectangular steps with different heights). So, a necessity in development of clear algorithm of calculation of all amplitudes in problems with the barriers of complicated shape has been appeared.

In~\cite{Maydanyuk.2000.UPJ} exact solution of the amplitudes of the stationary wave function for the barrier, located in the whole axis and composed of two rectangular steps of arbitrary shapes, was found. However, there was a question in determination of wave packets under the barrier on the first step (this was resolved in~\cite{Maydanyuk.2003.PhD-thesis}, part~1) and it was uncleared how to generalize such a formalism after complication of the barrier shape.
Some late, in~\cite{Esposito.2003.PRE} approach of multiple internal reflections of the waves was considered for tunneling of the particle through a number of equal rectangular steps located in whole axis and separated on equal distances one from another. However, solutions of the amplitudes, the coefficients of penetrability and reflections were presented for two such steps only, in a approximation when they are separated on enough large distance. Besides, the found solutions in approach of multiple internal reflections are constructed on the basis of the amplitudes of total wave function, obtained early by standard direct method (see Appendix A, (7), (18), (19) in~\cite{Esposito.2003.PRE}). Therefore, a question about convergence of sums of the amplitudes of the multiple waves by approach MIR with the corresponding amplitudes of total wave function by standard direct method inside each spacial region (and, especially, inside the barrier region) remains open.
So, \emph{we come to a serious problem of realization of the approach of multiple reflections in real quantum systems with barriers} (solution of which can call in question \underline{effectiveness} of applicability of the approach MIR, at all).

After complication of the barrier shape the second question has been appeared:
\emph{Whether is interference between separate waves (packets) appeared, which are formed after separate reflections relatively \underline{different boundaries} and reduces (or increases) the total wave? Whether does this come to principally different results of the approach of multiple internal reflections and direct method of quantum mechanics? Note that such interference cannot be appeared in the problem of tunneling through one rectangular barrier (and motion above it) and, therefore, in previous papers it was not visible.} This problem has already been calling in question a \underline{correctness} of approach MIR in application to the barriers with the complicated shapes.



The given paper is directed on study of such questions and gives answer on them.
In Sec.~2 the formalism of the method MIR in description of tunneling of the particle under one-dimensional rectangular barrier and its motion above it in whole axis (based on \cite{Maydanyuk.2000.UPJ,Maydanyuk.2002.JPS}, \cite{Maydanyuk.2003.PhD-thesis} part~1) is presented. Here, we put proof of the method, analyze its peculiarities, write solutions for amplitudes (without phase times).
Further, we present \underline{exact solutions} of the amplitudes, the coefficients of penetrability and reflection by the method MIR in description of tunneling of the particle through arbitrary number of potential rectangular steps with arbitrary heights and widths in full consideration of the multiple internal reflections relatively all boundaries (at first time, a main idea was introduced in~\cite{Maydanyuk.2003.PhD-thesis}, see p.~127--129). According to analysis, the total amplitudes of transmission and reflection inside each region and concerning the total barrier coincide with the corresponding amplitudes obtained by the standard method. \emph{Such formalism transforms the method of MIR, early used mainly in problems with one-dimensional rectangular barriers, into a power tool of calculation of \underline{exact solutions} of WF, the coefficients of penetrability and reflection relatively one-dimensional potentials of arbitrary shape, used in description of real quantum systems}.

In Sec.~3 we present the formalism of MIR in study of tunneling processes in problems of scattering of a particle on nucleus and $\alpha$-decay of the nucleus in the spherically symmetric approximation. At first, both processes are studied concerning a radial barrier of the simplest shape -- as rectangular step. In the scattering problem the packets described evolution of this process ($n$-multiple WP formed in result of successive $n$ transmissions and reflections relatively the boundaries of the barrier; total WPs inside each region; total WP transmitted through the barrier and total reflected WP from it), phase time of tunneling of the particle through the barrier and phase time of its reflection from the barrier are found. Expression for S-matrix is presented as a sum of two components, which are defined by the amplitudes of the stationary parts of the transmitted packet through the barrier and the reflected packet from it and correspond to resonant and potential scattering of the particle on the nucleus.

Using idea in~\cite{Maydanyuk.2002.PAST}, a main attention in this paper we concentrate to development of the formalism of MIR in description of the $\alpha$-decay. At first, the $\alpha$-decay with a radial rectangular barrier is considered. Here, the packet transmitted through the barrier which describes propagation of the particle from the internal region outside after its tunneling through the barrier, and the packet reflected from the barrier which describes delay of the particle inside the internal region in result of its reflection from the barrier are found; non-stationary analysis is fulfilled where \emph{duration of $\alpha$-decay of nucleus} is defined and calculated (see~\cite{Maydanyuk.2003.PhD-thesis}, part 2).
Further, in Sec.~5 we at first time present the formalism of MIR in description of the $\alpha$-decay with a realistic $\alpha$-nucleus radial barrier, when this barrier is approximated by arbitrary number of rectangular potential steps. In such approach, we demonstrate convergence of found solutions for the amplitudes of transmission and reflection for leaving of the $\alpha$-particle relatively such multi-steps potential with barrier, tending such potential to the initial realistic $\alpha$-nucleus one. Limit values of the amplitudes and coefficients are considered as exact concerning the realistic $\alpha$-nucleus potential (without application of approximation WKB), on the basis of which we define half-live. In approach MIR a dependence of the penetrability coefficient and half-live on location of starting point $R_{\rm start}$ is observed, from where the $\alpha$-particle begins to move outside. On this basis, we define a radial coordinate of the most probable formation of the $\alpha$-particle inside nucleus before its $\alpha$-decay as such starting point, at leaving of the $\alpha$-particle from which the half-live is maximally closed to its experimental value. It turns out, that half-lives calculated by such approach for a number of nuclei with the same $Z=84$ and different mass number $A$, are closer essentially to their experimental values in a comparison with half-lives found in~\cite{Buck.1993.ADNDT} by approach of WKB, that demonstrates effectiveness of the method MIR in description of the $\alpha$-decay with the realistic $\alpha$-nucleus barriers and in calculation of half-lives.

\section{Tunneling of the particle through barrier in one-dimensional problem
\label{sec.2}}

At first, we shall describe a general formalism of multiple internal reflections, using a simplest problem of tunneling of the particle through one-dimensional rectangular barrier in whole axis.

\subsection{Tunneling of wave through one rectangular barrier in whole axis
\label{sec.2.1}}

\begin{figure}[htbp]
\centerline{\includegraphics[width=50mm]{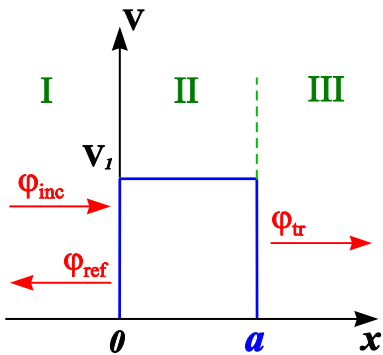}}
\caption{\small
Tunneling of the particle through one-dimensional rectangular barrier
\label{fig.2.1}}
\end{figure}

Let's consider a problem of tunneling of a particle in a positive $x$-direction through an one-dimensional rectangular potential barrier (see Fig.~\ref{fig.2.1}). Let's label a region I for $x < 0$, a region II for $0 < x < a$ and a region III for $x > a$, accordingly. Let's study an evolution of its tunneling through the barrier.


In a beginning we consider a standard approach to a solution of this problem \cite{Landau.v3.1989,Babicov.1988}. Let's consider a case when levels of energy lay under than a height of the barrier: $E < V_{1}$.
The tunneling evolution of the particle can be described using a non-stationary consideration of a propagating WP
\begin{equation}
  \psi(x, t) = \int\limits_{0}^{+\infty} g(E - \bar{E}) \varphi(k, x)
                e^{-iEt/\hbar} dE,
\label{eq.2.1.1}
\end{equation}
where the stationary WF has a form:
\begin{equation}
\varphi(x) = \left\{
\begin{array}{ll}
   e^{ikx}+A_{R}e^{-ikx},               & \mbox{for } x<0;   \\
   \alpha e^{\xi x} + \beta e^{-\xi x}, & \mbox{for } 0<x<a; \\
   A_{T} e^{ikx},                       & \mbox{for } x>a;
\end{array} \right.
\label{eq.2.1.2}
\end{equation}
and $k   = \frac{1}{\hbar}\sqrt{2mE}$,
    $\xi = \frac{1}{\hbar}\sqrt{2m(V_{1}-E)}$,
$E$ and $m$ are the total energy and mass of the particle, accordingly.
The weight amplitude $g(E - \bar{E})$ can be written in a form of gaussian \cite{Olkhovsky.1992.PRPLC} and satisfies to a requirement of the normalization $\int |g(E - \bar{E})|^{2} dE = 1$, value $\bar{E}$ is an average energy of the particle. One can calculate coefficients $A_{T}$, $A_{R}$, $\alpha$ and $\beta$ analytically, using a requirements of a continuity of WF $\varphi(x)$ and its derivative on each boundary of the barrier.

Substituting in Eq.~(\ref{eq.2.1.1}) instead of $\varphi(k, x)$ the incident $\varphi_{inc}(k, x)$, transmitted  $\varphi_{tr}(k, x)$ or reflected part of WF $\varphi_{ref}(k, x)$, defined by Eq.~(\ref{eq.2.1.2}), we receive the incident, transmitted  or reflected WP, accordingly.

We assume, that a time, for which the WP tunnels through the barrier, is enough small. So, the time necessary for a tunneling of an $\alpha$-particle through a barrier of decay in $\alpha$-decay of a nucleus, is about $10^{-21}$ seconds. We consider, that one can neglect a spreading of the WP for this time. And a breadth of the WP appears essentially more narrow on a comparison with a barrier breadth \cite{Olkhovsky.1992.PRPLC,Olkhovsky.1997.Trieste}.
Considering only sub-barrier processes, we exclude a component of waves for above-barrier energies, having included the additional transformation
\begin{equation}
   g(E - \bar{E}) \to g(E - \bar{E}) \theta(V_{1} - E),
\label{eq.2.1.3}
\end{equation}
where $\theta$-function satisfies to the requirement
%
\[
\theta(\eta) = \left\{
\begin{array}{ll}
   0,               & \mbox{for } \eta<0;   \\
   1,               & \mbox{for } \eta>0.
\end{array} \right.
\]

The method of multiple internal reflections considers the propagation process of the WP describing a motion of the particle, sequentially on steps of its penetration in relation to each boundary of the barrier
\cite{Fermor.1966.AJPIA,McVoy.1967.RMPHA,Anderson.1989.AJPIA}. Using this method, we find expressions for the transmitted and reflected WP in relation to the barrier.

At the first step we consider the WP in the region I, which is incident upon the first (initial) boundary of the barrier. Let's assume, that this package transforms into the WP, transmitted through this boundary and tunneling further in the region II, and into the WP, reflected from the boundary and propagating back in the region I. Thus we consider, that the WP, tunneling in the region II, is not reached the second (final) boundary of the barrier because of a terminating velocity of its propagation, and consequently at this step we consider only two regions I and II. Because of physical reasons to construct an expression for this packet, we consider, that its amplitude should decrease in a positive $x$-direction. We use only one item $\beta\exp(-\xi x)$ in Eq.~(\ref{eq.2.1.2}), throwing the second increasing item $\alpha\exp(\xi x)$ (in an opposite case we break a requirement of a finiteness of the WF for an indefinitely wide barrier). In result, in the region II we obtain:
\begin{equation}
  \psi^{1}_{tr}(x, t) = \int\limits_{0}^{+\infty} g(E - \bar{E})
  \theta(V_{1} - E) \beta^{0} e^{-\xi x -iEt/\hbar} dE,
  \mbox{for } 0<x<a.
\label{eq.2.1.4}
\end{equation}
Thus the WF in the barrier region constructed by such way, is an analytic continuation of a relevant expression for the WF, corresponding to a similar problem with above-barrier energies, where as a stationary expression we select the wave $\exp(ik_{2}x)$, propagated to the right.

Let's consider the first step further. One can write expressions for the incident and the reflected WP in relation to the first boundary as follows
\begin{equation}
\begin{array}{lcll}
\psi_{inc}(x, t) & = & \int\limits_{0}^{+\infty} g(E - \bar{E})
        \theta(V_{1} - E) e^{ikx -iEt/\hbar} dE,
        & \mbox{for } x<0, \\
\psi^{1}_{ref}(x, t) & = & \int\limits_{0}^{+\infty} g(E - \bar{E})
        \theta(V_{1} - E) A_{R}^{0} e^{-ikx -iEt/\hbar} dE,
        & \mbox{for } x<0.
\end{array}
\label{eq.2.1.5}
\end{equation}
A sum of these expressions represents the complete WF in the region I, which is dependent on a time. Let's require, that this WF and its derivative continuously transform into the WF (\ref{eq.2.1.4}) and its derivative at point $x=0$
(we assume, that the weight amplitude $g(E - \bar{E})$ differs weakly at transmitting and reflecting of the WP in relation to the barrier boundaries). In result, we obtain two equations, in which one can pass from the time-dependent WP to the corresponding stationary WF and obtain the unknown coefficients $\beta^{0}$ and $A_{R}^{0}$.

At the second step we consider the WP, tunneling in the region II and incident upon the second boundary of the barrier at point $x = a$. It transforms into the WP, transmitted through this boundary and propagated in the region III, and into the WP, reflected from the boundary and tunneled back in the region II. For a determination of these packets one can use Eq.~(\ref{eq.2.1.1}) with account (\ref{eq.2.1.3}), where as the stationary WF we use:
\begin{equation}
\begin{array}{lcll}
\varphi_{inc}^{2}(k, x) & = & \varphi_{tr}^{1}(k, x) =
        \beta^{0} e^{-\xi x},
        & \mbox{for } 0<x<a, \\
\varphi_{tr}^{2}(k, x) & = & A_{T}^{0}e^{ikx},
        & \mbox{for } x>a, \\
\varphi_{ref}^{2}(k, x) & = & \alpha^{0} e^{\xi x},
        & \mbox{for } 0<x<a.
\end{array}
\label{eq.2.1.6}
\end{equation}
Here, for forming an expression for the WP reflected from the boundary, we select an increasing part of the stationary solution $\alpha^{0} \exp(\xi x)$ only. Imposing a condition of continuity on the time-dependent WF and its derivative at point $x = a$, we obtain 2 new equations, from which we find the unknowns coefficients $A_{T}^{0}$ and $\alpha^{0}$.

At the third step the WP, tunneling in the region II, is incident upon the first boundary of the barrier. Then it transforms into the WP, transmitted through this boundary and propagated further in the region I, and into the WP, reflected from boundary and tunneled back in the region II. For a determination of these packets one can use Eq.~ (\ref{eq.2.1.1}) with account Eq.~(\ref{eq.2.1.3}), where as the stationary WF we use:
\begin{equation}
\begin{array}{lcll}
\varphi_{inc}^{3}(k, x) & = & \varphi_{ref}^{2}(k, x),
        & \mbox{for } 0<x<a, \\
\varphi_{tr}^{3}(k, x) & = & A_{R}^{1}e^{-ikx},
        & \mbox{for } x<0, \\
\varphi_{ref}^{3}(k, x) & = & \beta^{1} e^{-\xi x},
        & \mbox{for } 0<x<a.
\end{array}
\label{eq.2.1.7}
\end{equation}
Using a conditions of continuity for the time-dependent WF and its derivative at point $x = 0$, we obtain the unknowns coefficients $A_{R}^{1}$ and $\beta^{1}$.

Analyzing further possible processes of the transmission (and the reflection) of the WP through the boundaries of the barrier, we come to a deduction, that any of following steps can be reduced to one of 2 considered above. For the unknown coefficients $\alpha^{n}$, $\beta^{n}$,$A_{T}^{n}$ and $A_{R}^{n}$, used in expressions for the WP, forming in result of some internal reflections from the boundaries, one can obtain the recurrence relations:
\begin{equation}
\begin{array}{lll}
\beta^{0} = \displaystyle\frac{2k}{k+i\xi},     &
\alpha^{n} = \beta^{n} \displaystyle\frac{i\xi-k}{i\xi+k}e^{-2\xi a}, &
\beta^{n+1} = \alpha^{n} \displaystyle\frac{i\xi-k}{i\xi+k}, \\
A_{R}^{0} = \displaystyle\frac{k-i\xi}{k+i\xi},     &
A_{T}^{n} = \beta^{n} \displaystyle\frac{2i\xi}{i\xi+k}e^{-\xi a-ika}, &
A_{R}^{n+1} = \alpha^{n} \displaystyle\frac{2i\xi}{i\xi+k}.
\end{array}
\label{eq.2.1.8}
\end{equation}

Considering the propagation of the WP by such way, we obtain expressions for the WF on each region which can be written through series of multiple WP. Using Eq.~(\ref{eq.2.1.1}) with account Eq.~(\ref{eq.2.1.3}), we determine resultant expressions for the incident, transmitted and reflected WP in relation to the barrier, where one can need to use following expressions for the stationary WF:
\begin{equation}
\begin{array}{lcll}
\varphi_{inc}(k, x) & = & e^{ikx},
                        & \mbox{for } x<0, \\
\varphi_{tr}(k, x)  & = & \sum\limits_{n=0}^{+\infty} A_{T}^{n} e^{ikx},
                        & \mbox{for } x>a, \\
\varphi_{ref}(k, x) & = & \sum\limits_{n=0}^{+\infty} A_{R}^{n} e^{-ikx},
                        & \mbox{for } x<0.
\end{array}
\label{eq.2.1.9}
\end{equation}

Now we consider the WP formed in result of sequential $n$ reflections from the boundaries of the barrier and incident upon one of these boundaries at point $x = 0$ ($i = 1$) or at point $x = a$ ($i = 2$). In result, this WP transforms into the WP $\psi_{tr}^{i}(x, t)$, transmitted through boundary with number $i$, and into the WP $\psi_{ref}^{i}(x, t)$, reflected from this boundary. For an independent on $x$ parts of the stationary WF one can write:
\begin{equation}
\begin{array}{ll}
   \displaystyle\frac{\varphi_{tr}^{1}}{\exp(-\xi x)} =
   T_{1}^{+} \displaystyle\frac{\varphi_{inc}^{1}}{\exp(ikx)}, &
   \displaystyle\frac{\varphi_{ref}^{1}}{\exp(-ikx)} =
   R_{1}^{+} \displaystyle\frac{\varphi_{inc}^{1}}{\exp(ikx)}, \\
   \displaystyle\frac{\varphi_{tr}^{2}}{\exp(ikx)} =
   T_{2}^{+} \displaystyle\frac{\varphi_{inc}^{2}}{\exp(-\xi x)}, &
   \displaystyle\frac{\varphi_{ref}^{2}}{\exp(\xi x)} =
   R_{2}^{+} \displaystyle\frac{\varphi_{inc}^{2}}{\exp(-\xi x)}, \\
   \displaystyle\frac{\varphi_{tr}^{1}}{\exp(-ikx)} =
   T_{1}^{-} \displaystyle\frac{\varphi_{inc}^{1}}{\exp(\xi x)}, &
   \displaystyle\frac{\varphi_{ref}^{1}}{\exp(-\xi x)} =
   R_{1}^{-} \displaystyle\frac{\varphi_{inc}^{1}}{\exp(\xi x)},
\end{array}
\label{eq.2.1.10}
\end{equation}
where the sign ``+'' (or ``-'') corresponds to the WP, tunneling (or propagating) in a positive (or negative) $x$-direction and incident upon the boundary with number $i$. Using $T_{i}^{\pm}$ and $R_{i}^{\pm}$, one can precisely describe an arbitrary WP which has formed in result of $n$-multiple reflections, if to know a ``path'' of its propagation along the barrier. Using the recurrence relations Eq.~(\ref{eq.2.1.8}), the coefficients $T_{i}^{\pm}$ and $R_{i}^{\pm}$ can be obtained.
\begin{equation}
\begin{array}{lll}
T_{1}^{+} = \beta^{0},
&
T_{2}^{+} = \displaystyle\frac{A_{T}^{n}}{\beta^{n}},
&
T_{1}^{-} = \displaystyle\frac{A_{R}^{n+1}}{\alpha^{n}},
\\
R_{1}^{+} = A_{R}^{0},
&
R_{2}^{+} = \displaystyle\frac{\alpha^{n}}{\beta^{n}},
&
R_{1}^{-} = \displaystyle\frac{\beta^{n+1}}{\alpha^{n}}.
\end{array}
\label{eq.2.1.11}
\end{equation}

Using the recurrence relations, one can find series of coefficients $\alpha^{n}$, $\beta^{n}$, $A_{T}^{n}$ and $A_{R}^{n}$. However, these series can be calculated easier, using coefficients $T_{i}^{\pm}$ and $R_{i}^{\pm}$. Analyzing all possible ``paths'' of the WP propagations along the barrier, we receive:
\begin{equation}
\begin{array}{lcl}
\sum\limits_{n=0}^{+\infty} A_{T}^{n} & = &
        T_{2}^{+}T_{1}^{-} \biggl(1 + \sum\limits_{n=1}^{+\infty}
                                 (R_{2}^{+}R_{1}^{-})^{n} \biggr) =
        \displaystyle\frac{i4k \xi e^{-\xi a-ika}}{F_{sub}},  \\
\sum\limits_{n=0}^{+\infty} A_{R}^{n} & = &
        R_{1}^{+} + T_{1}^{+}R_{2}^{+}T_{1}^{-} \biggl(1 +
        \sum\limits_{n=1}^{+\infty}(R_{2}^{+}R_{1}^{-})^{n} \biggr) =
        \displaystyle\frac{k_{0}^{2}D_{-}}{F_{sub}},         \\
\sum\limits_{n=0}^{+\infty} \alpha^{n} & = &
        \alpha^{0} \biggl(1 + \sum\limits_{n=1}^{+\infty}
                           (R_{2}^{+}R_{1}^{-})^{n} \biggr) =
        \displaystyle\frac{2k(i\xi - k)e^{-2\xi a}}{F_{sub}}, \\
\sum\limits_{n=0}^{+\infty} \beta^{n} & = &
        \beta^{0} \biggl(1 + \sum\limits_{i=1}^{+\infty}
                          (R_{2}^{+}R_{1}^{-})^{n} \biggr) =
        \displaystyle\frac{2k(i\xi + k)}{F_{sub}}, \\
\end{array}
\label{eq.2.1.12}
\end{equation}
where
\begin{equation}
\begin{array}{lll}
F_{sub} & = & (k^{2} - \xi^{2})D_{-} + 2ik\xi D_{+},            \\
D_{\pm}   & = & 1 \pm e^{-2\xi a},                              \\
k_{0}^{2} & = & k^{2} + \xi^{2} = \displaystyle\frac{2mV_{1}}{\hbar^{2}}.
\end{array}
\label{eq.2.1.13}
\end{equation}

All series $\sum \alpha^{n}$, $\sum \beta^{n}$, $\sum A_{T}^{n}$ and $\sum A_{R}^{n}$, obtained using the method of multiple internal reflections, coincide with the corresponding coefficients $\alpha$, $\beta$, $A_{T}$ and $A_{R}$ of the Eq.~(\ref{eq.2.1.2}), calculated by a stationary methods \cite{Olkhovsky.1992.PRPLC,Landau.v3.1989,Razavy.1988.PRPLC}. Using the following substitution
\begin{equation}
  i\xi \to k_{2},
\label{eq.2.1.14}
\end{equation}
where $k_{2}= \frac{1}{\hbar}\sqrt{2m (E-V_{1})}$ is a wave number for a case of above-barrier energies, expression for the coefficients $\alpha^{n}$, $\beta^{n}$, $A_{T}^{n}$ and $A_{R}^{n}$ for each step, expressions for the WF for each step, the total Eqs.~(\ref{eq.2.1.12}) and (\ref{eq.2.1.13}) transform into the corresponding expressions for a problem of the particle propagation above this barrier. At the transformation of the WP and the time-dependent WF one can need to change a sign of argument at $\theta$-function. Besides the following property is fulfilled:
\begin{equation}
  \biggl|\sum\limits_{n=0}^{+\infty} A_{T}^{n}\biggr|^{2} +
  \biggl|\sum\limits_{n=0}^{+\infty} A_{R}^{n}\biggr|^{2} = 1.
\label{eq.2.1.15}
\end{equation}
\subsection{Tunneling of the particle through arbitrary number of rectangular steps with arbitrary heights and widths
\label{sec.2.2}}

Now let's consider propagation of the particle (from the left) through potential composed of arbitrary number of rectangular steps with arbitrary heights and widths. At first, we analyze a case of motion of the particle above the barrier. We have the following stationary wave function:
\begin{equation}
\varphi(x) = \left\{
\begin{array}{lll}
   e^{ikx} + A_{R}\,e^{-ikx},                           & \mbox{at } x \leq x_{1},             & \mbox{(region 1)}; \\
   \alpha_{2}\, e^{ik_{2}x} + \beta_{2}\, e^{-ik_{2}x}, & \mbox{at } x_{1} \leq x \leq x_{2},  & \mbox{(region 2)}; \\
   \alpha_{3}\, e^{ik_{3}x} + \beta_{3}\, e^{-ik_{3}x}, & \mbox{at } x_{2} \leq x \leq x_{3},  & \mbox{(region 3)}; \\
   \ldots & \ldots & \ldots \\
   \alpha_{N-1}\, e^{ik_{N-1}x} + \beta_{N-1}\, e^{-ik_{N-1}x}, & \mbox{at } x_{N-2} \leq x \leq x_{N-1},  & \mbox{(region N-1)}; \\
   A_{T}\,e^{ikx},                                      & \mbox{at } x \geq x_{N-1},           & \mbox{(region N)}
\end{array} \right.
\label{eq.2.2.1}
\end{equation}
where $\alpha_{j}$ and $\beta_{j}$ are unknown coefficients, $A_{T}$ and $A_{R}$ are unknown amplitudes of transmission and reflection.

The method of multiple internal reflections allows to describe propagation of the particle in such field with barriers and to find the transmitted and reflected waves. According to the method, propagation of the particle through the barrier is considered by use of wave packet successively by steps of its propagation relatively each boundary of the barrier. We conclude that any further step in such consideration will be similar to one from the first independent $2N-1$ steps. From the analysis of these steps we combine recurrent relations for calculation of amplitudes $A^{(n)}$, $S^{(n)}$, $\alpha^{(n)}$ and $\beta^{(n)}$ for arbitrary step with number $n$ and determine total amplitudes in each region. Further, we shall present results of calculation of amplitudes by approach MIR.

At first, we determine coefficients $T_{1}^{\pm}$, $T_{2}^{\pm}$ \ldots $T_{N-1}^{\pm}$ and $R_{1}^{\pm}$, $R_{2}^{\pm}$ \ldots $R_{N-1}^{\pm}$:
\begin{equation}
\begin{array}{ll}
\vspace{2mm}
   T_{j}^{+} = \displaystyle\frac{2k_{j}}{k_{j}+k_{j+1}} \,e^{i(k_{j}-k_{j+1}) x_{j}}, &
   T_{j}^{-} = \displaystyle\frac{2k_{j+1}}{k_{j}+k_{j+1}} \,e^{i(k_{j}-k_{j+1}) x_{j}}, \\
   R_{j}^{+} = \displaystyle\frac{k_{j}-k_{j+1}}{k_{j}+k_{j+1}} \,e^{2ik_{j}x_{j}}, &
   R_{j}^{-} = \displaystyle\frac{k_{j+1}-k_{j}}{k_{j}+k_{j+1}} \,e^{-2ik_{j+1}x_{j}}.
\end{array}
\label{eq.2.2.2}
\end{equation}
We find the following recurrent relations:
\begin{equation}
\begin{array}{l}
   \vspace{1mm}
   \tilde{R}_{j-1}^{+} =
     R_{j-1}^{+} + T_{j-1}^{+} \tilde{R}_{j}^{+} T_{j-1}^{-}
     \Bigl(1 + \sum\limits_{m=1}^{+\infty} (\tilde{R}_{j}^{+}R_{j-1}^{-})^{m} \Bigr) =
     R_{j-1}^{+} +
     \displaystyle\frac{T_{j-1}^{+} \tilde{R}_{j}^{+} T_{j-1}^{-}} {1 - \tilde{R}_{j}^{+} R_{j-1}^{-}}, \\

   \vspace{1mm}
   \tilde{R}_{j+1}^{-} =
     R_{j+1}^{-} + T_{j+1}^{-} \tilde{R}_{j}^{-} T_{j+1}^{+}
     \Bigl(1 + \sum\limits_{m=1}^{+\infty} (R_{j+1}^{+} \tilde{R}_{j}^{-})^{m} \Bigr) =
     R_{j+1}^{-} +
     \displaystyle\frac{T_{j+1}^{-} \tilde{R}_{j}^{-} T_{j+1}^{+}} {1 - R_{j+1}^{+} \tilde{R}_{j}^{-}}, \\

   \tilde{T}_{j+1}^{+} =
     \tilde{T}_{j}^{+} T_{j+1}^{+}
     \Bigl(1 + \sum\limits_{m=1}^{+\infty} (R_{j+1}^{+} \tilde{R}_{j}^{-})^{m} \Bigr) =
     \displaystyle\frac{\tilde{T}_{j}^{+} T_{j+1}^{+}} {1 - R_{j+1}^{+} \tilde{R}_{j}^{-}},
\end{array}
\label{eq.2.2.3}
\end{equation}
and select the following coefficients as starting:
\begin{equation}
\begin{array}{ccc}
  \tilde{R}_{N-1}^{+} = R_{N-1}^{+}, &
  \tilde{R}_{1}^{-} = R_{1}^{-}, &
  \tilde{T}_{1}^{+} = T_{1}^{+}.
\end{array}
\label{eq.2.2.4}
\end{equation}
Using them, we calculate successively the coefficients $\tilde{R}_{N-2}^{+}$ \ldots $\tilde{R}_{1}^{+}$, $\tilde{R}_{2}^{-}$ \ldots $\tilde{R}_{N-1}^{-}$ and $\tilde{T}_{2}^{+}$ \ldots $\tilde{T}_{N-1}^{+}$.

Now we determine coefficients $\beta_{j}$:
\begin{equation}
\begin{array}{l}
   \vspace{1mm}
   \beta_{j} =
     \tilde{T}_{j-1}^{+}
     \Bigl(1 + \sum\limits_{m=1}^{+\infty} (\tilde{R}_{j}^{+} \tilde{R}_{j-1}^{-})^{m} \Bigr) =
     \displaystyle\frac{\tilde{T}_{j-1}^{+}} {1 - \tilde{R}_{j}^{+} \tilde{R}_{j-1}^{-}}, \\
\end{array}
\label{eq.2.2.5}
\end{equation}
and amplitudes of transmission and reflection:
\begin{equation}
\begin{array}{cc}
  A_{T} = \tilde{T}_{N-1}^{+}, &
  A_{R} = \tilde{R}_{1}^{+}.
\end{array}
\label{eq.2.2.6}
\end{equation}
After comparison of the amplitudes of wave function obtained by such way with results obtained by standard direct method of quantum mechanics for the potential, composed of two rectangular steps of arbitrary height we obtain exact coincidence. Increasing of number of steps confirms coincidence of such results.
In particular, we check a property:
\begin{equation}
  |A_{T}|^{2} + |A_{R}|^{2} = 1.
\label{eq.2.2.7}
\end{equation}
All this sure us in that the presented above method MIR gives exact solutions of all amplitudes of wave function in all region of its definition.

Now we consider a case when energy of the particle is less then the height of one step with number $m$. Then for description of propagation of packet inside such step we must use such transformation:
\begin{equation}
  k_{m} \to i\,\xi_{m}.
\label{eq.2.2.8}
\end{equation}
In this case, we obtain also full coincidence between all amplitudes calculated by method MIR and by standard direct method. So, we generalize the method MIR for description of tunneling of the particle through multi-steps potential with arbitrary number of steps higher then the particle energy.

\section{Tunneling of the particle through a spherically symmetric barriers
\label{sec.3}}

The one-dimensional formalism of the method of multiple internal reflections in Sec.~\ref{sec.2} can be generalized for description of a motion of a particle (with its possible tunneling) in spherically symmetric field with barrier, allowing to study evolution of nuclear collisions and $\alpha$-decay of nuclei in spherically symmetric consideration.

\subsection{Scattering of the particle in central field with radial rectangular barrier
\label{sec.3.1}}

\begin{figure}[htbp]
\centerline{\includegraphics[width=50mm]{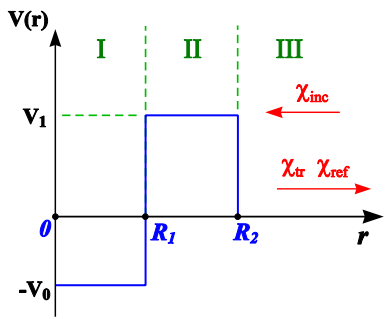}}
\caption{\small
Scattering of the particle on the radial rectangular barrier with its tunneling
\label{fig.3.1}}
\end{figure}

A problem of scattering of the particle on nucleus in spherically symmetric approximation is reduced to the problem of scattering of the particle with reduced mass inside the spherically symmetric field with barrier \cite{Maydanyuk.JPS.2002}. We shall present spherically symmetric formalism of multiple internal reflections for the problem with barrier of the simplest form (see Fig.~\ref{fig.3.1}):
\begin{equation}
  V(r) = \left\{
  \begin{array}{rll}
    -V_{0}, & \mbox{at } r<R_{1};       & \mbox{(region I)};  \\
     V_{1}, & \mbox{at } R_{1}<r<R_{2}, & \mbox{(region II)}; \\
     0,     & \mbox{at } r>R_{2},       & \mbox{(region III)}.
  \end{array} \right.
\label{eq.3.1.1}
\end{equation}
We note the region I for $0<r<R_{1}$, region II for $R_{1}<r<R_{2}$, and region III for $r>R_{2}$. We assume that the particle being under the action of central force $V(r)$ is incident on the external boundary $R_{2}$ outside. We shall consider a case when orbital moment is $l = 0$ and total energy $E$ is less then the barrier height $V_{1}$: $E<V_{1}$.

\subsubsection{Transmitted, reflected packets and S-matrix
\label{sec.3.1.1}}

At first, we describe standard approach for description of evolution of this process. Solving stationary Schr\"{o}dinger equation in each region, we find wave function:
\begin{equation}
  \psi(r, \theta, \varphi) =  \frac{\chi(r)}{r} Y_{lm}(\theta, \varphi),
\label{eq.3.1.1.1}
\end{equation}
\begin{equation}
  \chi(r) = \left\{
  \begin{array}{lll}
    A(e^{-ik_{1}r} - e^{ik_{1}r}),       & \mbox{at }  r<R_{1},      & \mbox{(region I)}, \\
    \alpha e^{\xi r} + \beta e^{-\xi r}, & \mbox{at } R_{1}<r<R_{2}, & \mbox{(region II)}, \\
    e^{-ikr} + Se^{ikr},                 & \mbox{at } r>R_{2},       & \mbox{(region III)}
  \end{array} \right.
\label{eq.3.1.1.2}
\end{equation}
where $Y_{lm}(\theta, \varphi)$ is spherical function,
    $k = \frac{1}{\hbar}\sqrt{2mE}$,
    $k_{1} = \frac{1}{\hbar}\sqrt{2m(E+V_{0})}$,
    $\xi = \frac{1}{\hbar}\sqrt{2m(V_{1}-E)}$
(here, we use boundary condition of the radial WF at $r=0$ to be finite, and its part describing incident of the particle on the barrier in region III is normalized on 1). The unknown coefficients $S$, $A$, $\alpha$ and $\beta$ are determined from condition of continuity of WF and its derivative at each boundary $R_{1}$ or $R_{2}$.

The standard approach to description of evolution of scattering of the particle on barrier consists in use of non-stationary wave packets, constructed on the basis of stationary components $\chi(k, r)$. Taking into account only sub-barrier energies, we obtain:
\begin{equation}
  \chi(r, t) = \int\limits_{0}^{+\infty} g(E - \bar{E}) \Theta(V_{l} - E) \chi(k, r) e^{-iEt/\hbar} dE,
\label{eq.3.1.1.3}
\end{equation}
\begin{equation}
  V_{l}(r) = V(r) + \displaystyle\frac{\hbar^{2}}{2m} \displaystyle\frac{l(l+1)}{r^{2}},
\label{eq.3.1.1.4}
\end{equation}
where the second item in (\ref{eq.3.1.1.4}) represent centrifugal energy, which equals to zero at $l = 0$, and weight amplitude $g(E - \bar{E})$ and average energy $\bar{E}$ are defined like one-dimensional case.

The problem of a motion of a particle in spherically symmetric field is reduced to the radial problem. The radial Schrodinger equation and normalization condition for the radial WF $\psi(r) = \chi(r) / r$ is reduced to one-dimensional Schr\"{o}dinger equation with centrifugal potential and the normalization condition for $\chi(r)$ \cite{Landau.v3.1989}. We describe the particle incident on the external boundary in region III by plane wave $\exp(-ikr)$? and scattering particle by the barrier in region III --- by outgoing plane wave $S \exp(ikr)$, which include both wave, described reflection of particle from the barrier, and wave, described transmission of the particle through the barrier when it tunnels into region I and then propagates back to region III. The reflected and transmitted waves are formed only one component $S \exp(ikr)$ and it is unclear how to separate them.

Being non-stationary, the method of multiple internal reflections allows to study the process of scattering of this particle inside the spherically symmetric field with the barrier in more details.
According to the method, the scattering of the particle on the barrier is considered by use of non-stationary WP consequently by steps of its propagation relatively each boundary of the barrier (like one-dimensional problem). In result of analysis we conclude that any step of further propagations of multiple WPs will be like one from only the first 4 independent steps. Analyzing these steps, we combine recurrent relations for calculation of the coefficients $A^{(n)}$, $S^{(n)}$, $\alpha^{(n)}$ and $\beta^{(n)}$ for arbitrary step with number $n$.

The total non-stationary WF in each region, taking into account multiple reflections of WPs relatively boundaries, can be presented in form of series of these WPs. Analyzing all possible ``paths'' of propagation of packets and using formalism of $T^{\pm}_{i}$ and $R^{\pm}_{i}$, we find sums:
\begin{equation}
\begin{array}{lcl}
\sum\limits_{n=1}^{+\infty} S^{(n)} & = &
     \displaystyle\frac{1}{F_{sub}}
     T_{2}^{-}T_{2}^{+} (R_{1}^{-}(1-R_{1}^{+}R_{0}^{-}) +
     T_{1}^{-}R_{0}^{-}T_{1}^{+}) = \\
 & = & \displaystyle\frac{4ik\xi
     \biggl(\displaystyle\frac{i\xi-k_{1}}{i\xi+k_{1}} -
     e^{2ik_{1}R_{1}}\biggr) e^{2\xi(R_{1}-R_{2}) - 2ikR_{2}} }
     {F_{sub} (k+i\xi)^{2}}, \\
\sum\limits_{n=0}^{+\infty}A^{(n)} & = &
    \displaystyle\frac{T_{1}^{-}T_{2}^{-}}{F_{sub}} =
    \displaystyle\frac{4ik\xi e^{-ikR_{2} + ik_{1}R_{1} -
    \xi(R_{2}-R_{1})}} {F_{sub} (k+i\xi)(k_{1}+i\xi)}, \\
\sum\limits_{n=0}^{+\infty}\alpha^{(n)} & = &
    \alpha^{(0)} \displaystyle\frac{1-R_{1}^{+}R_{0}^{-}}{F_{sub}} =
    \displaystyle\frac{2k \biggl(1 + \displaystyle\frac{k_{1}-i\xi}
    {k_{1}+i\xi}e^{2ik_{1}R_{1}}\biggr) e^{-(\xi+ik)R_{2}}}
    {F_{sub} (k+i\xi)}, \\
\sum\limits_{n=0}^{+\infty}\beta^{(n)} & = &
    \displaystyle\frac{\sum\limits_{n=0}^{+\infty}\alpha^{(n)} -
    T_{2}^{-}}{R_{2}^{+}} =
    \alpha^{(0)} \displaystyle\frac{R_{1}^{-}(1 - R_{1}^{+}R_{0}^{-})
    + T_{1}^{-}R_{0}^{-}T_{1}^{+}}{F_{sub}} = \\
 & = & \displaystyle\frac{2k
     \biggl(\displaystyle\frac{i\xi-k_{1}}{i\xi+k_{1}} -
     e^{2ik_{1}R_{1}}\biggr) e^{\xi(2R_{1}-R_{2}) - ikR_{2}} }
     {F_{sub} (k+i\xi) },
\end{array}
\label{eq.3.1.1.5}
\end{equation}
where
\begin{equation}
\begin{array}{lcl}
F_{sub} & = & (1 - R_{1}^{+}R_{0}^{-})(1 - R_{2}^{+}R_{1}^{-}) -
        R_{2}^{+}T_{1}^{-}R_{0}^{-}T_{1}^{+} =
        1 + \displaystyle\frac{k_{1}-i\xi}{k_{1}+i\xi}
        e^{2ik_{1}R_{1}} - \\
& - &   \displaystyle\frac{(k-i\xi)(k_{1}-i\xi)}{(k+i\xi)(k_{1}+i\xi)}
        e^{-2\xi (R_{2} - R_{1})} -
        \displaystyle\frac{k-i\xi}{k+i\xi}
        e^{-2\xi (R_{2}-R_{1}) + 2ik_{1}R_{1}}.
\end{array}
\label{eq.3.1.1.6}
\end{equation}
\begin{equation}
\begin{array}{ll}
T_{2}^{-} = \alpha^{(0)} =
            \displaystyle\frac{2k}{k+i\xi} e^{-(\xi+ik)R_{2}}, &
R_{2}^{-} = S^{(0)} =
            \displaystyle\frac{-i\xi+k}{i\xi+k}e^{-2ikR_{2}}, \\
T_{1}^{-} = \displaystyle\frac{A^{(n)}}{\alpha^{(n)}} =
            \displaystyle\frac{2i\xi}{i\xi+k_{1}} e^{(\xi+ik_{1})R_{1}}, &
R_{1}^{-} = \displaystyle\frac{\beta^{(n)}}{\alpha^{(n)}} =
            \displaystyle\frac{i\xi-k_{1}}{i\xi+k_{1}} e^{2\xi R_{1}}, \\
T_{0}^{-} = 0, &
R_{0}^{-} = 1, \\
T_{1}^{+} = \displaystyle\frac{\beta^{(n+1)}}{A^{(n)}} = -
            \displaystyle\frac{2k_{1}}{i\xi+k_{1}} e^{(\xi+ik_{1})R_{1}}, &
R_{1}^{+} = \displaystyle\frac{A^{(n+1)}}{A^{(n)}} =
            \displaystyle\frac{i\xi-k_{1}}{i\xi+k_{1}} e^{2ik_{1}R_{1}}, \\
T_{2}^{+} = \displaystyle\frac{S^{(n+1)}}{\beta^{(n)}} =
            \displaystyle\frac{2i\xi}{i\xi+k} e^{-(\xi+ik)R_{2}}, &
R_{2}^{+} = \displaystyle\frac{\alpha^{(n+1)}}{\beta^{(n)}} =
            \displaystyle\frac{i\xi-k}{i\xi+k} e^{-2\xi R_{2}}.
\end{array}
\label{eq.3.1.1.7}
\end{equation}
Here, the coefficients $T_{i}^{\pm}$ и $R_{i}^{\pm}$ are defined relatively boundary with number $i$ ($i=0$ for $r=0$, $i=1$ for $r=R_{1}$ and $i=2$ for $r=R_{2}$) and are calculated by use of recurrent expressions between $S^{(n)}$, $A^{(n)}$, $\alpha^{(n)}$ and $\beta^{(n)}$.

Now we define incident, transmitted and reflected WPs relatively the barrier (in region III):
\begin{equation}
\begin{array}{lcl}
\chi_{inc}(r, t) & = & \int\limits_{0}^{+\infty} g(E - \bar{E})
        \Theta(V_{1} - E) e^{-ikr -iEt/\hbar} dE, \\
\chi_{tr}(r, t)  & = & \int\limits_{0}^{+\infty} g(E - \bar{E})
        \Theta(V_{1} - E) S_{tr} e^{ikr -iEt/\hbar} dE, \\
\chi_{ref}(r, t) & = & \int\limits_{0}^{+\infty} g(E - \bar{E})
        \Theta(V_{1} - E) S_{ref} e^{ikr -iEt/\hbar} dE,
\end{array}
\label{eq.3.1.1.8}
\end{equation}
where
\begin{equation}
\begin{array}{rcl}
S       = S_{tr} + S_{ref}, &
S_{tr}  = \sum\limits_{n=1}^{+\infty}S^{(n)}, &
S_{ref} = S^{(0)}.
\end{array}
\label{eq.3.1.1.9}
\end{equation}
So, method MIR separates S-matrix into two components, corresponding to the amplitudes of the transmitted and reflected WPs relatively the barrier. This property fulfills for any energy level and allows to find coefficients of penetrability and reflection of the particle relatively the barrier (and also probabilities of processes of scattering through compound nucleus formation and potential scattering).

The coefficients $S^{(n)}$, $A^{(n)}$, $\alpha^{(n)}$ and $\beta^{(n)}$ for each step, WFs for each step, the coefficients $T_{i}^{\pm}$ and $R_{i}^{\pm}$, total sums of the coefficients $S^{(n)}$, $A^{(n)}$, $\alpha^{(n)}$ and $\beta^{(n)}$ at (\ref{eq.2.1.14}) transform into corresponding expressions for motion of the particle above the barrier. The total sums (\ref{eq.3.1.1.5}) of the coefficients $S^{(n)}$, $A^{(n)}$, $\alpha^{(n)}$ and $\beta^{(n)}$ coincide with corresponding coefficients $S$, $A$, $\alpha$ and $\beta$, calculated in the direct method of QM. We have:
\begin{equation}
  |S| = 1.
\label{eq.3.1.1.10}
\end{equation}

\subsubsection{Phase times of tunneling and reflection relatively the barrier
\label{sec.3.1.2}}

According to method of stationary phase, we have
\begin{equation}
  \displaystyle\frac{\partial}{\partial E} \mbox{arg } \chi_{inc}(r, t) =
  \displaystyle\frac{\partial}{\partial E} \mbox{arg } \chi_{tr}(r, t)  =
  \displaystyle\frac{\partial}{\partial E} \mbox{arg } \chi_{ref}(r, t) =
  \mbox{const}.
\label{eq.3.1.2.1}
\end{equation}
Now we consider the first step. Let maximum of WP in region III be incident to the external boundary of the barrier at point $R_{2}$ at time moment $t_{inc}$. Using (\ref{eq.3.1.2.1}), we define time moment $t_{ref}^{(1)}$ of leaving of the maximum of the reflected WP outside in region III:
\begin{equation}
t_{ref}^{(1)} = t_{inc} +
        \displaystyle\frac{2mR_{2}}{\hbar k} +
        \hbar\displaystyle\frac{\partial \mbox{arg } S^{(0)}}{\partial E}.
\label{eq.3.1.2.2}
\end{equation}
For time moment $t_{tr}^{(n)}$ of leaving from the barrier region outside of the maximum of $n$-multiple WP we write:
\begin{equation}
t_{tr}^{(n)} = t_{inc} +
        \displaystyle\frac{2mR_{2}}{\hbar k} +
        \hbar\displaystyle\frac{\partial \mbox{arg } S^{(n)}}{\partial E}.
\label{eq.3.1.2.3}
\end{equation}
Using (\ref{eq.3.1.2.1}) at point $r = R_{2}$ and taking into account (\ref{eq.3.1.2.2}), (\ref{eq.3.1.2.3}), we find time of transmission of WP through the barrier and time of reflection of WP from the barrier:
\begin{equation}
\begin{array}{l}
\tau_{tr} = t_{tr} - t_{inc} =
        \displaystyle\frac{2mR_{2}}{\hbar k} +
        \hbar\displaystyle\frac{\partial \mbox{arg } S_{tr}}{\partial E}, \\
\tau_{ref} = t_{ref} - t_{inc} =
        \displaystyle\frac{2mR_{2}}{\hbar k} +
        \hbar\displaystyle\frac{\partial \mbox{arg } S_{ref}}{\partial E}.
\end{array}
\label{eq.3.1.2.4}
\end{equation}

For WP tunneling under the barrier we obtain:
\begin{equation}
\begin{array}{l}
\tau_{tun} = \hbar\displaystyle\frac{\partial}{\partial E}
        \mbox{arg } \displaystyle\frac{i\xi-k_{1}-(i\xi+k_{1})e^{2ik_{1}R_{1}}}
        {(i\xi+k)^{2}(i\xi+k_{1}) F_{sub}}, \\
\tau_{ref} = \displaystyle\frac{2m}{\hbar\xi k}.
\end{array}
\label{eq.3.1.2.5}
\end{equation}
For propagated WP above the barrier we obtain:
\begin{equation}
\begin{array}{l}
\tau_{prop} = \displaystyle\frac{2m(R_{2}-R_{1})}{\hbar k_{2}} +
        \hbar\displaystyle\frac{\partial}{\partial E}
        \mbox{arg } \displaystyle\frac{k_{2}-k_{1}-
(k_{2}+k_{1})e^{2ik_{1}R_{1}}}
        {(k+k_{2})(k_{1}+k_{2})F_{above}}, \\
\tau_{ref} = 0,
\end{array}
\label{eq.3.1.2.6}
\end{equation}
where $F_{above}$ can be obtained from $F_{sub}$ at (\ref{eq.2.1.14}). Also we have:
\begin{equation}
\begin{array}{l}
\tau_{tr} =
        \displaystyle\frac{2mR_{2}}{\hbar k} +
        \hbar\displaystyle\frac{\partial \mbox{arg } S}{\partial E} +
        \hbar\displaystyle\frac{\partial \mbox{arg } \gamma}{\partial E}, \\
\tau_{ref} =
        \displaystyle\frac{2mR_{2}}{\hbar k} +
        \hbar\displaystyle\frac{\partial \mbox{arg } S}{\partial E} +
        \hbar\displaystyle\frac{\partial \mbox{arg }
        (1-\gamma)}{\partial E},
\end{array}
\label{eq.3.1.2.7}
\end{equation}
where
\begin{equation}
  \gamma = \displaystyle\frac{S_{tr}}{S}.
\label{eq.3.1.2.8}
\end{equation}
One can see that tunneling and reflection times are different only by one item \cite{Maydanyuk.VANT.2002.p20}.

Now we consider scattering of the particle on enough high and wide barrier. In such case for sub-barrier tunneling we obtain:
\begin{equation}
\tau_{tun} = \displaystyle\frac{2m}{\hbar k \xi} +
        \displaystyle\frac{4mR_{1}\sin{2k_{1}R_{1}}(1-2\cos{2k_{1}R_{1}})}
        {\hbar \xi(1-\cos{2k_{1}R_{1}})}.
\label{eq.3.1.2.9}
\end{equation}
Here, the following consequence of approximations are used:
$\xi (R_{2} - R_{1}) \to +\infty$, $\xi \to +\infty$, $R_{2} - R_{1} \to +\infty$.

\subsection{$\alpha$-decay}
\label{sec.3.2}

Now let's generalize the method MIR for description of $\alpha$-decay in spherically symmetric approximation.
We shall assume that starting from some time moment (one can name it as \emph{time moment $t_{\rm formation}$ of formation of the $\alpha$-particle and daughter nucleus}) the decaying nucleus can be considered as composite system of the $\alpha$-particle and daughter nucleus. We come from a problem of motion of the $\alpha$-particle in the field of the daughter nucleus to another problem of propagation of the particle with reduced mass $m$ in spherical symmetric field with a radial barrier.

For simplicity, we shall consider a case when total energy of system $E$ is higher then the barrier height $V_{1}$: $E>V_{1}$.
For forming a main idea of application of the multiple internal reflections for description of $\alpha$-decay, we shall assume that the radial potential $V(r)$ can be separated into 3 regions: internal region I, region of the barrier II, and external region III,
\begin{figure}[htbp]
\centerline{\includegraphics[width=50mm]{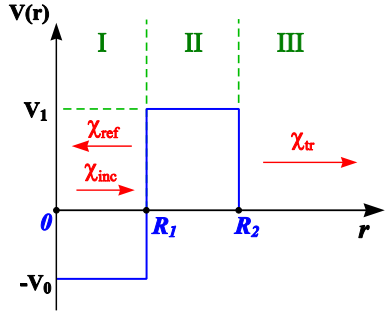}}
\caption{\small
Propagation of wave packet from the internal region outside with its tunneling through the barrier in the $\alpha$-decay
\label{fig.3.2}}
\end{figure}
and in each region we know exact general solution of the stationary wave function in the form:
\begin{equation}
  \psi(r, \theta, \varphi) =  \frac{\chi(r)}{r}\: Y_{lm}(\theta, \varphi),
\label{eq.3.2.1.1}
\end{equation}
\begin{equation}
  \chi(r) = \left\{
  \begin{array}{lll}
    A_{1} c_{1}^{-}(r) + B_{1} c_{1}^{+}(r), & \mbox{at } 0 \le r \le R_{1}     & \mbox{(region I)}, \\
    A_{2} c_{2}^{-}(r) + B_{2} c_{2}^{+}(r), & \mbox{at } R_{1} \le r \le R_{2} & \mbox{(region II)}, \\
    B_{3} c_{3}^{+}(r),                      & \mbox{at } r \ge R_{2}           & \mbox{(region III)}
  \end{array} \right.
\label{eq.3.2.1.2}
\end{equation}
where $c_{i}^{-}(r)$ and $c_{i}^{+}(r)$ are ingoing and outgoing waves, $A_{i}$ and $B_{i}$ are constant coefficients. According to ideology of multiple internal reflections, we describe the $\alpha$-decay by use of wave packet describing the particle with reduced mass, successively by steps of its propagation from the internal region I outside through the barrier region II relatively each boundary (see Fig.~\ref{fig.3.2}).

In the first step, we consider beginning of the $\alpha$-decay as propagation of the packet in the region I outside at $t > t_{\rm formation}$, which at some time moment $t_{inc}$ is incident on the internal boundary of the barrier at $R_{1}$. We name this time moment as \emph{time moment of beginning of the $\alpha$-decay}.
The incident WP forms new WP in the region II at time moment $t_{tr}$ propagating from the boundary $R_{1}$ outside, and new WP in the region I at time moment $t_{ref}$ propagating from this boundary inside back. We define the incident, transmitted and reflected packets concerning the boundary $R_{1}$ so:
\begin{equation}
\begin{array}{lcl}
  \chi_{inc}(r, t) & = & 
    \int\limits_{0}^{+\infty} g(E - \bar{E}) \Theta(E - V_{1}) \chi_{inc}(r, k) e^{-iEt/\hbar} dE, \\
  \chi_{tr}^{(1)}(r, t)  & = & 
    \int\limits_{0}^{+\infty} g(E - \bar{E}) \Theta(E - V_{1}) \chi_{tr}^{(1)}(r, k) e^{-iEt/\hbar} dE, \\
  \chi_{ref}^{(1)}(r, t) & = & 
    \int\limits_{0}^{+\infty} g(E - \bar{E}) \Theta(E - V_{1}) \chi_{ref}^{(1)}(r, k) e^{-iEt/\hbar} dE
\end{array}
\label{eq.3.2.1.3}
\end{equation}
where $g(E - \bar{E})$ is gaussian for above-barrier energies and we have:
\begin{equation}
\begin{array}{ll}
  \chi_{inc}(r, k)       = B_{1}^{(1)} c_{1}^{+}(r), &  \mbox{at } 0 \le r \le R_{1}, \\
  \chi_{tr}^{(1)}(r, k)  = B_{2}^{(1)} c_{2}^{+}(r), &  \mbox{at } R_{1} \le r \le R_{2}, \\
  \chi_{ref}^{(1)}(r, k) = A_{1}^{(1)} c_{1}^{-}(r), &  \mbox{at } 0 \le r \le R_{1}
\end{array}
\label{eq.3.2.1.4}
\end{equation}
Also we shall neglect by variation of weight amplitude
\begin{equation}
  g_{inc}(E) \simeq g_{tr}^{(1)}(E) \simeq g_{ref}^{(1)}(E).
\label{eq.3.2.1.5}
\end{equation}

Using continuity condition for total wave function composed on these WPs and its derivative at $R_{1}$, we find the unknown coefficients $A_{1}^{(1)}$ and $B_{2}^{(1)}$:
\begin{equation}
\begin{array}{l}
  A_{1}^{(1)} = B_{1}^{(1)} \displaystyle\frac
  {c_{1}^{+}(r) \displaystyle\frac{\partial}{\partial r} c_{2}^{+}(r) -
  c_{2}^{+}(r) \displaystyle\frac{\partial}{\partial r} c_{1}^{+}(r)}
  {c_{2}^{+}(r) \displaystyle\frac{\partial}{\partial r} c_{1}^{-}(r) -
  c_{1}^{-}(r) \displaystyle\frac{\partial}{\partial r} c_{2}^{+}(r)}, \\
  B_{2}^{(1)} = B_{1}^{(1)} \displaystyle\frac
  {c_{1}^{+}(r) \displaystyle\frac{\partial}{\partial r} c_{1}^{-}(r) -
  c_{1}^{-}(r) \displaystyle\frac{\partial}{\partial r} c_{1}^{+}(r)}
  {c_{2}^{+}(r) \displaystyle\frac{\partial}{\partial r} c_{1}^{-}(r) -
  c_{1}^{-}(r) \displaystyle\frac{\partial}{\partial r} c_{2}^{+}(r)}.
\end{array}
\label{eq.3.2.1.6}
\end{equation}

Then, like one-dimensional problem (see Sec.~\ref{sec.2.1}) and the scattering problem of the particle on nucleus (see Sec.~\ref{sec.3.1}), we consider further propagation of WP through the barrier outside in next steps. Here, any nex step with propagation and reflection of WPs concerning the barrier boundaries can be reduced to one from 4 independent steps. For determination of the unknown $A_{i}^{(n)}$ and $B_{i}^{(n)}$ one can find recurrent relations and calculate the coefficients $T_{i}^{\pm}$ and $R_{i}^{\pm}$:

\begin{equation}
\begin{array}{ll}
  T_{1}^{+} = \displaystyle\frac{B_{2}^{(1)}}{B_{1}^{(1)}}, &
  T_{2}^{+} = \displaystyle\frac
    {c_{2}^{+}(R_{2})
    \displaystyle\frac{\partial}{\partial r} c_{2}^{-}(R_{2}) -
    c_{2}^{-}(R_{2})
    \displaystyle\frac{\partial}{\partial r} c_{2}^{+}(R_{2})}
    {c_{3}^{+}(R_{2})
    \displaystyle\frac{\partial}{\partial r} c_{2}^{-}(R_{2}) -
    c_{2}^{-}(R_{2})
    \displaystyle\frac{\partial}{\partial r} c_{3}^{+}(R_{2})}, \\
  R_{1}^{+} = \displaystyle\frac{A_{1}^{(1)}}{B_{1}^{(1)}}, &
  R_{2}^{+} = \displaystyle\frac
    {c_{2}^{+}(R_{2})
    \displaystyle\frac{\partial}{\partial r} c_{3}^{+}(R_{2}) -
    c_{3}^{+}(R_{2})
    \displaystyle\frac{\partial}{\partial r} c_{2}^{+}(R_{2})}
    {c_{3}^{+}(R_{2})
    \displaystyle\frac{\partial}{\partial r} c_{2}^{-}(R_{2}) -
    c_{2}^{-}(R_{2})
    \displaystyle\frac{\partial}{\partial r} c_{3}^{+}(R_{2})}, \\
  T_{0}^{-} = 0,  &
  T_{1}^{-} = \displaystyle\frac
    {c_{2}^{+}(R_{1})
    \displaystyle\frac{\partial}{\partial r} c_{2}^{-}(R_{1}) -
    c_{2}^{-}(R_{1})
    \displaystyle\frac{\partial}{\partial r} c_{2}^{+}(R_{1})}
    {c_{2}^{+}(R_{1})
    \displaystyle\frac{\partial}{\partial r} c_{1}^{-}(R_{1}) -
    c_{1}^{-}(R_{1})
    \displaystyle\frac{\partial}{\partial r} c_{2}^{+}(R_{1})}, \\
  R_{0}^{-} = -1, &
  R_{1}^{-} = \displaystyle\frac
    {c_{1}^{-}(R_{1})
    \displaystyle\frac{\partial}{\partial r} c_{2}^{-}(R_{1}) -
    c_{2}^{-}(R_{1})
    \displaystyle\frac{\partial}{\partial r} c_{1}^{-}(R_{1})}
    {c_{2}^{+}(R_{1})
    \displaystyle\frac{\partial}{\partial r} c_{1}^{-}(R_{1}) -
    c_{1}^{-}(R_{1})
    \displaystyle\frac{\partial}{\partial r} c_{2}^{+}(R_{1})}.
\end{array}
\label{eq.3.2.1.7}
\end{equation}
One can find total transmitted and reflected WPs relatively the barrier:

\begin{equation}
\begin{array}{lcl}
\chi_{tr}(r, t)  & = & \int\limits_{0}^{+\infty} g(E - \bar{E})
        \Theta(E - V_{1}) \sum_{n=1}^{+\infty}
        B_{3}^{(n)} c_{3}^{+}(r) e^{ikr -iEt/\hbar} dE, \\
\chi_{ref}(r, t) & = & \int\limits_{0}^{+\infty} g(E - \bar{E})
        \Theta(E - V_{1}) \sum_{n=1}^{+\infty}
        A_{1}^{(n)} c_{1}^{-}(r) e^{-ikr -iEt/\hbar} dE.
\end{array}
\label{eq.3.2.1.8}
\end{equation}
All sums of the coefficients $A_{i}^{(n)}$ and $B_{i}^{(n)}$ are calculated on the basis of $T_{i}^{\pm}$ and $R_{i}^{\pm}$. We obtain:
\begin{equation}
\begin{array}{l}
\sum\limits_{n=1}^{+\infty} A_{1}^{(n)} =
        B_{1}^{1} \displaystyle\frac{1}
        {R_{0}^{-} (1 - \widetilde{R}_{1}^{+} R_{0}^{-})}, \\
\sum\limits_{n=2}^{+\infty} B_{1}^{(n)} =
        B_{1}^{1} \displaystyle\frac{1}
        {1 - \widetilde{R}_{1}^{+} R_{0}^{-}}, \\
\sum\limits_{n=1}^{+\infty} A_{2}^{(n)} =
        B_{1}^{1} \displaystyle\frac{T_{1}^{+}}
        {\widetilde{R}_{1}^{-}(1 - R_{1}^{+} R_{0}^{-})
        (1 - \widetilde{R}_{1}^{-} R_{2}^{+})}, \\
\sum\limits_{n=1}^{+\infty} B_{2}^{(n)} =
        B_{1}^{1} \displaystyle\frac{T_{1}^{+}}
        {(1 - R_{1}^{+} R_{0}^{-})
        (1 - \widetilde{R}_{1}^{-} R_{2}^{+})}, \\
\sum\limits_{n=1}^{+\infty} B_{3}^{(n)} =
        B_{1}^{1} \displaystyle\frac{T_{1}^{+} T_{2}^{+}}
        {(1 - R_{1}^{+} R_{0}^{-})
        (1 - \widetilde{R}_{1}^{-} R_{2}^{+})}
\end{array}
\label{eq.3.2.1.9}
\end{equation}
where
\begin{equation}
\begin{array}{ll}
\widetilde{R}_{1}^{+} = R_{1}^{+} +
        \displaystyle\frac{T_{1}^{+} T_{1}^{-} R_{2}^{+}}
        {1 - R_{1}^{-} R_{2}^{+}}, &
\widetilde{R}_{1}^{+} = R_{1}^{-} +
        \displaystyle\frac{T_{1}^{+} T_{1}^{-} R_{0}^{-}}
        {1 - R_{1}^{+} R_{0}^{-}}.
\end{array}
\label{eq.3.2.1.10}
\end{equation}
All expressions for sums $\sum A_{i}^{(n)}$ and $\sum B_{i}^{(n)}$ obtained by method MIR coincide with the corresponding coefficients $A_{i}$ and $B_{i}$ calculated by standard direct method (see also \cite{Baz}).
If the total energy $E$ is less then the barrier height $V_{1}$, then expressions for sums of the coefficients $A_{i}^{(n)}$ and $B_{i}^{(n)}$ are obtained by use of (\ref{eq.3.2.1.6}), (\ref{eq.3.2.1.7}), (\ref{eq.3.2.1.9}), (\ref{eq.3.2.1.10}) and (\ref{eq.2.1.14}) for the barrier region II. These coefficients for arbitrary step are calculated on the basis of (\ref{eq.3.2.1.6}) and the coefficients $T_{i}^{\pm}$ and $R_{i}^{\pm}$. Amplitude $B_{1}^{(1)}$ is found from the normalization condition for the radial WF.

A non-stationary analysis of the $\alpha$-decay can be fulfilled, like one-dimensional problem (see~\cite{Maydanyuk.2002.JPS,Maydanyuk.2006.FPL}) or the scattering problem (see Sec.~\ref{sec.3.1}).
We shall estimate duration of the $\alpha$-decay on the basis of the barrier penetrability. In consideration of motion of the $\alpha$-particle from the internal region outside with its tunneling it needs to take into account multiple internal reflections inside the internal region I, formed in result of reflections of packets from the barrier inside. But, the barrier penetrability can be related to duration of decay and to multiple reflections inside the internal region by different ways. Therefore, we give two independent approaches for definition of the penetrability coefficient $T$ of the barrier, describing propagation of the particle from the internal region I into external region III.

\vspace{5mm}
\noindent
\textbf{\underline{Approach A:}}

In the first case, we define \emph{the coefficients of penetrability and reflection with taking into account of multiple internal reflections of packets inside the internal region}. Here, as the incident, transmitted and reflected stationary parts of ВФ we use $\chi_{inc}(r)$ from (\ref{eq.3.2.1.4}) and $\chi_{tr}(r)$, $\chi_{ref}(r)$ from (\ref{eq.3.2.1.8}). We have:
\begin{equation}
\begin{array}{l}
T = \displaystyle\frac{
  \biggl|\sum\limits_{n=2}^{+\infty} B_{3}^{(n)}\biggr|^{2}}
  {\biggl| B_{1}^{(1)}\biggr|^{2}} \displaystyle\frac{
  c_{3}^{+}(r)
  \displaystyle\frac{\partial}{\partial r} c_{3}^{+}(r)^{*} -
  c_{3}^{+}(r)^{*}
  \displaystyle\frac{\partial}{\partial r} c_{3}^{+}(r)}
  {c_{1}^{+}(r)
  \displaystyle\frac{\partial}{\partial r} c_{1}^{+}(r)^{*} -
  c_{1}^{+}(r)^{*}
  \displaystyle\frac{\partial}{\partial r} c_{1}^{+}(r)},      \\
R = \displaystyle\frac
  {\biggl|\sum\limits_{n=1}^{+\infty} A_{1}^{(n)}\biggr|^{2}}
  {\biggl| B_{1}^{(1)}\biggr|^{2}} \displaystyle\frac
  {c_{1}^{-}(r)
  \displaystyle\frac{\partial}{\partial r} c_{1}^{-}(r)^{*} -
  c_{1}^{-}(r)^{*}
  \displaystyle\frac{\partial}{\partial r} c_{1}^{-}(r)}
  {c_{1}^{+}(r)
  \displaystyle\frac{\partial}{\partial r} c_{1}^{+}(r)^{*} -
  c_{1}^{+}(r)^{*}
  \displaystyle\frac{\partial}{\partial r} c_{1}^{+}(r)}.
\end{array}
\label{eq.3.2.1.11}
\end{equation}
At such definition, the coefficients $T$ and $R$ are not separated evidently into factor, characterized the penetrability of the barrier directly without taking into account multiple reflections of packets inside the internal region I, and factor characterized such multiple internal reflections.

Here, we introduce \emph{duration of $\alpha$-decay} $\tau_{decay}$, defining it as difference between time moment $t_{tr}$ of leaving of the total transmitted WP from the barrier (\emph{time moment of finishing of $\alpha$-decay}) and time moment $t_{inc}$ of incident of WP inside region I on the internal boundary $R_{1}$ in the first step (\emph{time moment of beginning of $\alpha$-decay}). Also, we define \emph{time moment of reflection of the particle from the barrier} $\tau_{ref}$ as difference between time moment $t_{ref}$ of reflection of the total reflected WP from the barrier into the internal region I and $t_{inc}$. Using (\ref{eq.3.1.2.1}), we obtain:
\begin{equation}
\begin{array}{lcl}
\tau_{tr} & = & t_{tr} - t_{inc} = \hbar \biggl(
  \displaystyle\frac{\partial}{\partial E}\mbox{arg } c_{3}^{+}(R_{2}) -
  \displaystyle\frac{\partial}{\partial E}\mbox{arg } c_{1}^{+}(R_{1}) + \\
  & + &
  \displaystyle\frac{\partial}{\partial E}\mbox{arg }
  \biggl(\sum\limits_{n=1}^{+\infty} B_{3}^{(n)} \biggr) -
  \displaystyle\frac{\partial}{\partial E}\mbox{arg } B_{1}^{(1)} \biggr), \\
\tau_{ref} & = & t_{ref} - t_{inc} = \hbar \biggl(
  \displaystyle\frac{\partial}{\partial E}\mbox{arg } c_{1}^{-}(R_{1}) -
  \displaystyle\frac{\partial}{\partial E}\mbox{arg } c_{1}^{+}(R_{1}) + \\
  & + &
  \displaystyle\frac{\partial}{\partial E}\mbox{arg }
  \biggl(\sum\limits_{n=1}^{+\infty} A_{1}^{(n)} \biggr) -
  \displaystyle\frac{\partial}{\partial E}\mbox{arg } B_{1}^{(1)} \biggr).
\end{array}
\label{eq.3.2.1.12}
\end{equation}

\vspace{5mm}
\noindent
\textbf{\underline{Approach B:}}

In the second case, we define \emph{the penetrability coefficient of the barrier without taking into account of multiple internal reflections of packets inside the internal region}, i.~e. as in one-dimensional case. However, such internal reflections we include inside the internal region, like in the semiclassical way of definition of half-lives --- through introduction of independent normalized factor $F$ defined a number of ``collisions'' of the $\alpha$-particle with the barrier inside the internal region I. As non-stationary characteristic of the $\alpha$-decay, we define half-life $\tau$ on the basis of width $\Gamma$ of the $\alpha$-decay.

\subsubsection{Decay with radial $\alpha$-nucleus potential in a form of one rectangular barrier and rectangular well in the internal region
\label{sec.3.2.2}}

As an example of application of the method MIR to calculation of the barrier penetrability and estimation of time characteristics in approach A, we shall consider the $\alpha$-decay with the barrier of the simplest form:
\begin{equation}
  V(r) = \left\{
  \begin{array}{rll}
    -V_{0}, & \mbox{at } r<R_{1};       & \mbox{(region I)};  \\
     V_{1}, & \mbox{at } R_{1}<r<R_{2}, & \mbox{(region II)}; \\
     0,     & \mbox{at } r>R_{2},       & \mbox{(region III)}.
  \end{array} \right.
\label{eq.3.2.2.1}
\end{equation}
Let's orbital moment is $l=0$ and total energy $E$ less then the barrier height. Stationary WF is:
\begin{equation}
  \psi(r, \theta, \varphi) =  \frac{\chi(r)}{r} Y_{lm}(\theta, \varphi),
\label{eq.3.2.2.2}
\end{equation}
\begin{equation}
  \chi(r) = \left\{
  \begin{array}{lll}
    A(e^{-ik_{1}r} - e^{ik_{1}r}),       & \mbox{at }  r<R_{1},      & \mbox{(region I)}, \\
    \alpha e^{\xi r} + \beta e^{-\xi r}, & \mbox{at } R_{1}<r<R_{2}, & \mbox{(region II)}, \\
    Se^{ikr},                            & \mbox{at } r>R_{2},       & \mbox{(region III)}
  \end{array} \right.
\label{eq.3.2.2.3}
\end{equation}
where $Y_{lm}(\theta, \varphi)$ is spherical function,
    $k = \frac{1}{\hbar}\sqrt{2mE}$,
    $k_{1} = \frac{1}{\hbar}\sqrt{2m(E+V_{0})}$,
    $\xi = \frac{1}{\hbar}\sqrt{2m(V_{1}-E)}$.
Taking into account (\ref{eq.3.2.1.2}), we write:
\begin{equation}
\begin{array}{ll}
  c_{1}^{\pm} = e^{\pm ik_{1}r}, &
  c_{3}^{+} = e^{ikr}.
\end{array}
\label{eq.3.2.2.4}
\end{equation}

The coefficients $T_{i}^{\pm}$ and $R_{i}^{\pm}$ are calculated on the basis of (\ref{eq.3.2.1.7}) and using form of $c_{i}^{\pm}$:
\begin{equation}
\begin{array}{ll}
  T_{1}^{+} = \displaystyle\frac{2 k_{1}}{i\xi + k_{1}}
              e^{\xi R_{1} + ik_{1} R_{1}}, &
  R_{1}^{+} = \displaystyle\frac{k_{1} - i\xi}{k_{1} + i\xi}
              e^{2i k_{1} R_{1}}, \\
  T_{2}^{+} = \displaystyle\frac{2 i\xi}{k + i\xi}
              e^{-\xi R_{2} - ik R_{2}}, &
  R_{2}^{+} = \displaystyle\frac{i\xi - k}{i\xi + k}
              e^{-2 \xi R_{2}}, \\
  T_{0}^{-} = 0,  &
  R_{0}^{-} = -1, \\
  T_{1}^{-} = \displaystyle\frac{2 i\xi}{i\xi + k_{1}}
              e^{\xi R_{1} + ik_{1} R_{1}}, &
  R_{1}^{-} = \displaystyle\frac{i\xi - k_{1}}{i\xi + k_{1}}
              e^{2 \xi R_{1}}.
\end{array}
\label{eq.3.2.2.5}
\end{equation}

The coefficients of penetrability $T$ and reflection $R$ of the particle relatively the barrier can be obtained from (\ref{eq.3.2.1.11}). Time of tunneling (время распада ядра) and time of reflection can be found from (\ref{eq.3.2.1.12}). We obtain:
\begin{equation}
\begin{array}{ll}
  \tau_{tr}  = \hbar\displaystyle\frac{\partial}{\partial E}
               \arg{C_{tr}}, &
  \tau_{ref} = \hbar\displaystyle\frac{\partial}{\partial E}
               \arg{C_{ref}}.
\end{array}
\label{eq.3.2.2.6}
\end{equation}
where
\begin{equation}
\begin{array}{lcl}
C_{tr}  & = &
        \displaystyle\frac{i4k_{1}\xi}
        {(k+i\xi)(k_{1}+i\xi + (k_{1}-i\xi)exp(2ik_{1}R_{1}))} \times \\
        & \times &
        \Biggl(1 -
        \displaystyle\frac
        {(i\xi-k)(i\xi-k_{1} - (i\xi+k_{1})exp(2ik_{1}R_{1}))}
        {(i\xi+k)(i\xi+k_{1} - (i\xi-k_{1})exp(2ik_{1}R_{1}))}
        exp(2\xi(R_{1}-R_{2})\Biggr)^{-1};      \\
C_{ref} & = &
        \displaystyle\frac{C_{1}}{C_{2}};       \\
C_{1}   & = &
        \displaystyle\frac{i\xi-k_{1}}{i\xi+k_{1}}exp(2ikR_{1})
        \Biggl(1 -
        \displaystyle\frac
        {(i\xi-k)(i\xi-k_{1})}{(i\xi+k)(i\xi+k_{1})}
        exp(2\xi(R_{1}-R_{2}))\Biggr) +         \\
        & + &
        \displaystyle\frac
        {4ik\xi(k-i\xi)}{(i\xi+k)^{2}(i\xi+k_{1})}
        exp(2\xi(R_{1}-R_{2}) + 2ik_{1}R_{1});  \\
C_{2}   & = &
        \Biggl(1 +
        \displaystyle\frac{k_{1}-i\xi}{k_{1}+i\xi}exp(2ikR_{1})
        \Biggr)
        \Biggl(1 -
        \displaystyle\frac
        {(i\xi-k)(i\xi-k_{1})}{(i\xi+k)(i\xi+k_{1})}
        exp(2\xi(R_{1}-R_{2}))\Biggr) +         \\
        & + &
        \displaystyle\frac
        {4ik\xi(k-i\xi)}{(i\xi+k)^{2}(i\xi+k_{1})}
        exp(2\xi(R_{1}-R_{2}) + 2ik_{1}R_{1}).
\end{array}
\label{eq.3.2.2.7}
\end{equation}

\subsubsection{Decay with the radial $\alpha$-nucleus potential composed of arbitrary number of rectangular steps
\label{sec.3.2.3}}

Now we shall consider the $\alpha$-decay with a realistic $\alpha$-nucleus radial barrier, when this barrier is enough well approximated by number of rectangular potential steps. The first region I we define starting from point $R_{\rm min}$, and we assume that the $\alpha$-particle is formed here and moves outside. Supposing that $\alpha$-particle can be formed enough far from zero, we use non-zero value for $R_{\rm min}$. As before, we shall analyze the motion of this particle on the basis of multiple internal reflections of WP successively by steps of its propagation relatively all boundaries between steps in the multi-step radial barrier. In consideration of multiple internal reflections of multiple packets inside radial internal region starting from zero and in dynamical estimation of the $\alpha$-decay, we use the second approach B and shall find half-live of the $\alpha$-decay. Then comparing obtained results with half-lives obtained by the semiclassical approach, we shall estimate how the method MIR can be effective.

At first, we write total wave function:
\begin{equation}
\chi(r) = \left\{
\begin{array}{lll}
   e^{ikr} + A_{R}\,e^{-ikr},                           &
     \mbox{at } R_{\rm min} < r \leq r_{1},         & \mbox{(region 1)}; \\
   \alpha_{2}\, e^{ik_{2}r} + \beta_{2}\, e^{-ik_{2}r}, &
     \mbox{at } r_{1} \leq r \leq r_{2},  & \mbox{(region 2)}; \\
   \alpha_{3}\, e^{ik_{3}r} + \beta_{3}\, e^{-ik_{3}r}, &
     \mbox{at } r_{2} \leq r \leq r_{3},  & \mbox{(region 3)}; \\
   \ldots & \ldots & \ldots \\
   \alpha_{n-1}\, e^{ik_{N-1}r} + \beta_{N-1}\, e^{-ik_{N-1}r}, &
     \mbox{at } r_{N-2} \leq r \leq r_{N-1}, & \mbox{(region N-1)}; \\
   A_{T}\,e^{ikr}, &
     \mbox{at } r_{N-1} \leq r \leq R_{\rm max},             & \mbox{(region N)}
\end{array} \right.
\label{eq.3.2.3.1}
\end{equation}
where $\alpha_{j}$ and $\beta_{j}$ are unknown amplitudes, $A_{T}$ and $A_{R}$ are unknown amplitudes of transmission and reflection.

Like one-dimensional problem, we define coefficients $T_{1}^{\pm}$, $T_{2}^{\pm}$ \ldots $T_{N-1}^{\pm}$ and $R_{1}^{\pm}$, $R_{2}^{\pm}$ \ldots $R_{N-1}^{\pm}$:
\begin{equation}
\begin{array}{ll}
\vspace{2mm}
   T_{j}^{+} = \displaystyle\frac{2k_{j}}{k_{j}+k_{j+1}} \,e^{i(k_{j}-k_{j+1}) r_{j}}, &
   T_{j}^{-} = \displaystyle\frac{2k_{j+1}}{k_{j}+k_{j+1}} \,e^{i(k_{j}-k_{j+1}) r_{j}}, \\
   R_{j}^{+} = \displaystyle\frac{k_{j}-k_{j+1}}{k_{j}+k_{j+1}} \,e^{2ik_{j}r_{j}}, &
   R_{j}^{-} = \displaystyle\frac{k_{j+1}-k_{j}}{k_{j}+k_{j+1}} \,e^{-2ik_{j+1}r_{j}}.
\end{array}
\label{eq.3.2.3.2}
\end{equation}
Using recurrent relations:
\begin{equation}
\begin{array}{l}
   \vspace{1mm}
   \tilde{R}_{j-1}^{+} =
     R_{j-1}^{+} + T_{j-1}^{+} \tilde{R}_{j}^{+} T_{j-1}^{-}
     \Bigl(1 + \sum\limits_{m=1}^{+\infty} (\tilde{R}_{j}^{+}R_{j-1}^{-})^{m} \Bigr) =
     R_{j-1}^{+} +
     \displaystyle\frac{T_{j-1}^{+} \tilde{R}_{j}^{+} T_{j-1}^{-}} {1 - \tilde{R}_{j}^{+} R_{j-1}^{-}}, \\

   \vspace{1mm}
   \tilde{R}_{j+1}^{-} =
     R_{j+1}^{-} + T_{j+1}^{-} \tilde{R}_{j}^{-} T_{j+1}^{+}
     \Bigl(1 + \sum\limits_{m=1}^{+\infty} (R_{j+1}^{+} \tilde{R}_{j}^{-})^{m} \Bigr) =
     R_{j+1}^{-} +
     \displaystyle\frac{T_{j+1}^{-} \tilde{R}_{j}^{-} T_{j+1}^{+}} {1 - R_{j+1}^{+} \tilde{R}_{j}^{-}}, \\

   \tilde{T}_{j+1}^{+} =
     \tilde{T}_{j}^{+} T_{j+1}^{+}
     \Bigl(1 + \sum\limits_{m=1}^{+\infty} (R_{j+1}^{+} \tilde{R}_{j}^{-})^{m} \Bigr) =
     \displaystyle\frac{\tilde{T}_{j}^{+} T_{j+1}^{+}} {1 - R_{j+1}^{+} \tilde{R}_{j}^{-}},
\end{array}
\label{eq.3.2.3.3}
\end{equation}
and selecting as starting the following values:
\begin{equation}
\begin{array}{ccc}
  \tilde{R}_{N-1}^{+} = R_{N-1}^{+}, &
  \tilde{R}_{1}^{-} = R_{1}^{-}, &
  \tilde{T}_{1}^{+} = T_{1}^{+},
\end{array}
\label{eq.3.2.3.4}
\end{equation}
we calculate successively coefficients $\tilde{R}_{N-2}^{+}$ \ldots $\tilde{R}_{1}^{+}$, $\tilde{R}_{2}^{-}$ \ldots
$\tilde{R}_{N-1}^{-}$ and $\tilde{T}_{2}^{+}$ \ldots $\tilde{T}_{N-1}^{+}$.
At finishing, we determine coefficients $\beta_{j}$:
\begin{equation}
\begin{array}{l}
   \vspace{1mm}
   \beta_{j} =
     \tilde{T}_{j-1}^{+}
     \Bigl(1 + \sum\limits_{m=1}^{+\infty} (\tilde{R}_{j}^{+} \tilde{R}_{j-1}^{-})^{m} \Bigr) =
     \displaystyle\frac{\tilde{T}_{j-1}^{+}} {1 - \tilde{R}_{j}^{+} \tilde{R}_{j-1}^{-}},
\end{array}
\label{eq.3.2.3.5}
\end{equation}
the amplitudes of transmission and reflection:
\begin{equation}
\begin{array}{cc}
  A_{T} = \tilde{T}_{N-1}^{+}, &
  A_{R} = \tilde{R}_{1}^{+}
\end{array}
\label{eq.3.2.3.6}
\end{equation}
and corresponding coefficients of penetrability $T$ and reflection $R$:
\begin{equation}
\begin{array}{cc}
  T_{MIR} = \displaystyle\frac{k_{n}}{k_{1}}\; \bigl|A_{T}\bigr|^{2}, &
  R_{MIR} = \bigl|A_{R}\bigr|^{2}.
\end{array}
\label{eq.3.2.3.7}
\end{equation}
As test of the found solutions, we use the property:
\begin{equation}
\begin{array}{ccc}
  \displaystyle\frac{k_{n}}{k_{1}}\; |A_{T}|^{2} + |A_{R}|^{2} = 1 & \mbox{ or }&
  T_{MIR} + R_{MIR} = 1.
\end{array}
\label{eq.3.2.3.8}
\end{equation}

\subsubsection{Width $\Gamma$ and half-live
\label{sec.3.2.4}}

We define width $\Gamma$ of the $\alpha$-decay, in semiclassical approximation, by following the procedure of \emph{Gurvitz} and \emph{K\"{a}lbermann} \cite{Gurvitz.1987.PRL}:
\begin{equation}
  \Gamma = P_{\alpha}\, F\: \displaystyle\frac{\hbar^{2}}{4m}\; T
\label{eq.3.2.4.1}
\end{equation}
where $P_{\alpha}$ is the $\alpha$-particle preformation probability and $F$ is the normalization factor.
$T$ is the penetrability coefficient in propagation of the particle from the internal region outside with its tunneling through the barrier, which we shall calculate by approach MIR ar by approach WKB. In approach MIR this coefficient we define so:
\begin{equation}
  T_{WKB} =
  \exp\;
  \Biggl\{
    -2 \displaystyle\int\limits_{R_{2}}^{R_{3}}
    \sqrt{\displaystyle\frac{2m}{\hbar^{2}}\: \Bigl(Q - V(r)\Bigr)} \; dr
  \Biggr\}
\label{eq.3.2.4.2}
\end{equation}
where $R_{2}$ and $R_{3}$ are the second and third turning points.
According to~\cite{Buck.1993.ADNDT}, the \emph{normalization factor} $F$ is given by simplified way so:
\begin{equation}
  F_{2} =
  \Biggl\{\:
    \displaystyle\int\limits_{R_{1}}^{R_{2}}
    \displaystyle\frac{dr}{2k(r)}
  \Biggr\}^{-1}
\label{eq.3.2.4.3}
\end{equation}
or by improved way so:
\begin{equation}
  F_{1} =
  \Biggl\{\:
    \displaystyle\int\limits_{R_{1}}^{R_{2}}
    \displaystyle\frac{1}{k(r)}\;
    \cos^{2} \Biggl[\:\displaystyle\int\limits_{R_{1}}^{r} k(r')\; dr' - \displaystyle\frac{\pi}{4} \Biggr]\; dr
  \Biggr\}^{-1}.
\label{eq.3.2.4.4}
\end{equation}
The $\alpha$-decay half-live is related to the width $\Gamma$ by well known expression:
\begin{equation}
  T_{1/2} = \hbar\; \ln 2 / \Gamma.
\label{eq.3.2.4.5}
\end{equation}

\section{$\alpha$-nucleus potential
\label{sec.4}}

For description of interaction between $\alpha$-particle and the daughter nucleus we shall use the $\alpha$-nucleus potential in the general form:
\begin{equation}
  V (r, \theta, l, Q) = v_{C} (r, \theta) + v_{N} (r, \theta, Q) + v_{l} (r)
\label{eq.4.0.1}
\end{equation}
where $v_{C} (r, \theta)$, $v_{N} (r, \theta, Q)$ and $v_{l} (r)$ are Coulomb, nuclear and centrifugal components. There are different approaches in definition of parameters for these components. In this paper we shall use two variants.

\subsection{$\alpha$-nucleus potential of V.~Yu.~Denisov and H.~Ikezoe
\label{sec.4.1}}

In the first case, we shall use the approach proposed by V.~Yu.~Denisov and H.~Ikezoe in~\cite{Denisov.2005.PHRVA} (for simplicity, we shall name the potential (\ref{eq.4.0.1}) with parameters defined according to \cite{Denisov.2005.PHRVA} as \emph{potential in the form DI}). According to (6)--(10) in~\cite{Denisov.2005.PHRVA}, Coulomb $v_{C} (r, \theta)$, nuclear $v_{N} (r, \theta, Q)$ and centrifugal $v_{l} (r)$ components have the following form:
\begin{equation}
\begin{array}{l}
  v_{C} (r, \theta) =
  \left\{
  \begin{array}{ll}
    \displaystyle\frac{2 Z e^{2}} {r}
      \biggl(1 + \displaystyle\frac{3 R^{2}} {5 r^{2}} \beta_{2} Y_{20}(\theta) \biggr), &
      \mbox{for  } r \ge r_{m}, \\
    \displaystyle\frac{2 Z e^{2}} {r_{m}}
      \biggl\{
        \displaystyle\frac{3}{2} -
        \displaystyle\frac{r^{2}}{2r_{m}^{2}} +
        \displaystyle\frac{3 R^{2}} {5 r_{m}^{2}} \beta_{2} Y_{20}(\theta)
        \Bigl(2 - \displaystyle\frac{r^{3}}{r_{m}^{3}} \Bigr)
      \biggr\}, &
      \mbox{for  } r < r_{m},
  \end{array}
  \right.
\end{array}
\label{eq.4.1.1}
\end{equation}
\begin{equation}
\begin{array}{l}
  \vspace{0mm}
  v_{N} (r, \theta, Q) = \displaystyle\frac{V(A,Z,Q)} {1 + \exp{\displaystyle\frac{r-r_{m}(\theta)} {d}}}, \\
\end{array}
\label{eq.4.1.2}
\end{equation}
\begin{equation}
\begin{array}{l}
  v_{l} (r) = \displaystyle\frac{l\,(l+1)} {2mr^{2}}.
\end{array}
\label{eq.4.1.3}
\end{equation}
We define the parameters of the Coulomb and nuclear components as (see relations ~(14), (16)--(19) in~\cite{Denisov.2005.PHRVA}):
\begin{equation}
\begin{array}{rcl}
  V(A,Z,Q) & = & -(30.275 - 0.45838 \, Z/A^{1/3} + 58.270\,I - 0.24244 \, Q),
\end{array}
\label{eq.4.1.4}
\end{equation}
\begin{equation}
\begin{array}{rcl}
  R & = & R_{p}\:(1 + 3.0909/R_{p}^{2}) + 0.1243\,t, \\
\end{array}
\label{eq.4.1.5}
\end{equation}
\begin{equation}
\begin{array}{rcl}
  R_{p} & = & 1.24 \,A^{1/3} \: (1 + 1.646/A - 0.191\,I), \\
\end{array}
\label{eq.4.1.6}
\end{equation}
\begin{equation}
\begin{array}{rcl}
  t & = & I - 0.4 \, A/(A+200), \\
\end{array}
\label{eq.4.1.7}
\end{equation}
\begin{equation}
\begin{array}{rcl}
  d & = & 0.49290, \\
\end{array}
\label{eq.4.1.8}
\end{equation}
\begin{equation}
\begin{array}{rcl}
  I & = & (A-2Z) / A.
\end{array}
\label{eq.4.1.9}
\end{equation}
According to relations (21)--(22) in~\cite{Denisov.2005.PHRVA}, we also use:
\begin{equation}
\begin{array}{cclccl}
  r_{m}(\theta) & = & 1.5268 + R (\theta), &
  \hspace{3mm}
  R(\theta)     & = & R \: (1 + \beta_{2} Y_{20}(\theta) ).
\end{array}
\label{eq.4.1.10}
\end{equation}
Here, $A$ and $Z$ are the nucleon and proton numbers of the daughter nucleus, respectively; $Q$ is the $Q$-value, for the $\alpha$-decay, $R$ is the radius of the daughter nucleus, $V(A,Z,Q,\theta)$ is the strength of the nuclear component; $r_{m}$ is the effective radius of the nuclear component, $d$ is the parameter of the diffuseness; $Y_{20}(\theta)$ is the spherical harmonic function of the second order, $\theta$ is the angle between the direction of the leaving $\alpha$-particle and the axis of the axial symmetry of the daughter nucleus; $\beta_{2}$ is the parameter of the quadruple deformation of the daughter nucleus. 

In this paper we restrict ourselves by spherical symmetric approximation of $\alpha$-decay. Then the Coulomb and nucleus components are transformed into the following:
\begin{equation}
\begin{array}{l}
  v_{C} (r, \theta) =
  \left\{
  \begin{array}{ll}
    \displaystyle\frac{2 Z e^{2}} {r}, &
      \mbox{for  } r \ge r_{m}, \\
    \displaystyle\frac{Z e^{2}} {r_{m}}\;
      \biggl\{ 3 -  \displaystyle\frac{r^{2}}{r_{m}^{2}} \biggr\}, &
      \mbox{for  } r < r_{m},
  \end{array}
  \right.
\end{array}
\label{eq.4.1.11}
\end{equation}
\begin{equation}
  v_{N} (r, \theta, Q) = \displaystyle\frac{V(A,Z,Q)} {1 + \exp{\displaystyle\frac{r-r_{m}} {d}}}
\label{eq.4.1.12}
\end{equation}
where
\begin{equation}
  r_{m} = 1.5268 + R.
\label{eq.4.1.13}
\end{equation}

\subsection{$\alpha$-nucleus potential of B.~Buck, A.~C.~Merchant and S.~P.~Perez
\label{sec.4.2}}

In the second case, we shall use approach proposed by B.~Buck, A.~C.~Merchant and S.~P.~Perez in~\cite{Buck.1993.ADNDT} (we shall name the potential (\ref{eq.4.0.1}) with the parameters defined according to \cite{Buck.1993.ADNDT} as \emph{potential in the form BMP}). According to (1)--(3) in~\cite{Buck.1993.ADNDT}, the nuclear component is:
\begin{equation}
  v_{N} (r) = -V_{0} \; \displaystyle\frac{1 + \cosh{R_{BMP}/d}} {\cosh{r/d} + \cosh{R_{BMP}/d}},
\label{eq.4.2.1}
\end{equation}
Coulomb component is:
\begin{equation}
\begin{array}{l}
  v_{C} (r) =
  \left\{
  \begin{array}{ll}
   \displaystyle\frac{2 Z e^{2}} {r}, &
     \mbox{for  }  r \ge R_{BMP}, \\
   \displaystyle\frac{Z e^{2}} {r_{m}}  \biggl(3 - \displaystyle\frac{r^{2}} {R_{BMP}^{2}} \biggr), &
     \mbox{for  }  r \le R_{BMP}
  \end{array}
  \right.
\end{array}
\label{eq.4.2.2}
\end{equation}
and centrifugal component is:
\begin{equation}
  v_{l} (r) = \displaystyle\frac{\hbar^{2}} {2m} \displaystyle\frac{(l+\frac{1}{2})^{2}} {r^{2}}.
\label{eq.4.2.3}
\end{equation}
Values $V_{0}$, $R_{BMP}$ and $d$ for the different nuclei are presented in tables in~\cite{Buck.1993.ADNDT}.

\section{Analysis of the method MIR in $\alpha$-decay 
\label{sec.5}}

We shall study, how the presented above method of multiple internal reflections really works (how much it is effective, accurate, gives stable result) in the problem of determination of half-lives in $\alpha$-decay of heavy nuclei, which can be considered in enough good approximation as spherically symmetric. Today, there is a lot of well developed methods of calculations of half-lives of $\alpha$-decay, they are experimentally studied well. So, for method MIR analysis we have rich theoretical and experimental material in this problem.

We shall use two $\alpha$-active nuclei: $^{210}\mbox{\rm Po}$ and $^{214}\mbox{\rm Po}$. Such a choice we explain by they have small coefficient of quadruple deformation $\beta_{2}$ and in enough good approximation can be considered as spherical. They have very similar shapes of barriers and parameters of $\alpha$-decay, but their half-lives (and $Q$-values) are different essentially (according to~\cite{Buck.1993.ADNDT}: $\tau_{\rm exp} (^{210}\mbox{\rm Po}) = 1,2 \cdot 10^{7}$~sec and $\tau_{\rm exp} (^{214}\mbox{\rm Po}) = 1,6 \cdot 10^{-4}$~sec). We shall be interesting in how accurately the method MIR allows to calculate these values and to describe such difference (in a comparison with the semiclassical approach). Besides, we presently have numerical methods of calculations of wave functions with high accuracy tested for these nuclei (that is shown essentially in calculations of spectra of photon bremsstrahlung accompanying $\alpha$-decay, see~\cite{Maydanyuk.2003.PTP,Maydanyuk.2006.EPJA,Maydanyuk.2008.EPJA,Maydanyuk.2008.MPLA}) and computer programs.

At first, we shall consider $\alpha$-decay of $^{214}\mbox{\rm Po}$. We shall use $\alpha$-nucleus potential in the spherically symmetric BMP form. We shall study $\alpha$-decay on the basis of leaving of the particle with reduced mass from the internal region outside with its tunneling through the barrier inside the region of $r$ with internal boundary $R_{\rm min}$ (we assume it to be very close to zero) and the external boundary $R_{\rm max}$ (we assume it to be very large). To apply the method MIR for description of this process, we change the radial $\alpha$-nucleus potential inside the studied region into approximated one, which consists from finite number $N$ of rectangular potential steps. We define the region of formation of the particle inside the first interval with $R_{\rm min} \le r \le R_{1}$ (here we assume: $Q > V(r)$), from here, this particle starts to move outside. According to method MIR, we consider leaving of the particle on the basis of WP consequently by steps of its propagation relatively each boundary.
Using technique of the coefficients $T^{\pm}_{j}$ and $R^{\pm}_{j}$ in (\ref{eq.3.2.3.2})--(\ref{eq.3.2.3.4}), we calculate amplitudes $\beta_{j}$ in each interval by (\ref{eq.3.2.3.5}), total amplitudes of transmission $A_{T}$ and reflection $A_{R}$ by (\ref{eq.3.2.3.6}). The penetrability coefficient $T_{MIR}$, describing leaving of the particle from the internal region outside with its tunneling in the region from $R_{\rm min}$ up to $R_{\rm max}$, we calculate by (\ref{eq.3.2.3.7}). In this paper we restrict ourselves by definition (\ref{eq.3.2.4.3}) for the normalization coefficient $F_{2}$ in WKB-approach, we find width $\Gamma$ by (\ref{eq.3.2.4.1}) and half-live $\tau_{MIR}$ by (\ref{eq.3.2.4.5}).

Further, we define the penetrability coefficient $T_{WKB}$ in the WKB-approach by (\ref{eq.3.2.4.2}), on the basis of which we recalculate $\Gamma$-width and half-live $\tau_{WKB}$ by (\ref{eq.3.2.4.1}) and (\ref{eq.3.2.4.5}). Comparing the found half-lives (also amplitudes of transmission and reflection, penetrability coefficients) by two approaches, we analyze the effectiveness of the method MIR for the studied nucleus.

\subsection{Analysis of convergence of the method MIR in dependence on the number of intervals inside the studied region \label{sec.5.1}}

Let's analyze whether the method MIR gives unique values for the penetrability coefficient, amplitudes of transmission (in the last interval with number $N$) and reflection (in the first interval with number 1) after taking into account all needed multiple internal reflections relatively all boundaries, in dependence on the number $N$ of intervals inside the region from $R_{\rm min}$ to $R_{\rm max}$. According to~\cite{Buck.1993.ADNDT}, for $\alpha$-decay of $^{214}{\rm Po}$ we use $Q_{\alpha}=7,865$~MeV, $R_{\rm BMP}=7,417$~fm. The calculated penetrability coefficient $T$ with increasing of the number $N$ is presented in table~\ref{table.1}.
\begin{table}
\begin{center}
\begin{tabular}{|c|c|c|c|c|c|c|c|} \hline
  Number $N$ &
  \multicolumn{2}{|c|}{Transmission amplitude} &
  \multicolumn{2}{|c|}{Reflection amplitude} &
  \multicolumn{2}{|c|}{Penetrability} &
  Half-live
  \\
  \cline{2-7}
  of intervals &
  $\Re\, (A_{T})$ &
  $\Im\, (A_{T})$ &
  $\Re\, (A_{R})$ &
  $\Im\, (A_{R})$ &
  $T$ &
  $\Delta_{T}$ &
  $\tau_{1/2}$, sec
  \\ \hline
  1000 &  6,97E-10  & -1,200E-9   & -0,976   & 0,213  & 1,2946E-18 &     -      & 1,55E-4 \\
  2000 &  6,612E-10 & -1,148E-9   & -0,9683  & 0,249  & 1,1790E-18 & 0,1156E-18 & 1,62E-4 \\
  3000 &  6,504E-10 & -1,140E-9   & -0,9636  & 0,267  & 1,1580E-18 & 0,0210E-18 & 1,87E-4 \\
  4000 &  6,452E-10 & -1,139E-9   & -0,9608  & 0,277  & 1,1507E-18 & 0,0073E-18 & 1,94E-4 \\
  5000 &  6,422E-10 & -1,1384E-9  & -0,9590  & 0,2832 & 1,1473E-18 & 0,0034E-18 & 1,77E-4 \\
  6000 &  6,403E-10 & -1,13834E-9 & -0,9577  & 0,2874 & 1,1454E-18 & 0,0019E-18 & 1,78E-4 \\
  7000 &  6,389E-10 & -1,13838E-9 & -0,9568  & 0,2905 & 1,1443E-18 & 0,0011E-18 & 1,72E-4 \\
  8000 &  6,379E-10 & -1,13848E-9 & -0,95613 & 0,2929 & 1,1436E-18 & 0,0007E-18 & 1,66E-4 \\
  9000 &  6,371E-10 & -1,13859E-9 & -0,95556 & 0,2947 & 1,1431E-18 & 0,0005E-18 & 1,77E-4 \\
 10000 &  6,365E-10 & -1,13871E-9 & -0,95510 & 0,2962 & 1,1428E-18 & 0,0003E-18 & 1,79E-4 \\
  \hline
\end{tabular}
\end{center}
\caption{Dependence of the amplitudes of transmission and reflection, penetrability coefficient and half-live on number $N$ of intervals in the region from $R_{\rm min}$ up to $R_{\rm max}$ for $\alpha$-decay of $^{214}{\rm Po}$ (we use: $R_{\rm min}=0,11$~fm, $R_{\rm max}=100$~fm;
$\Delta_{T} = |T_{n+1} - T_{n}|$ is parameter of convergence in calculations of the penetrability coefficient $T$ where $T_{n}$ and $T_{n+1}$ are the penetrability coefficients for each previous and further numbers $N$;
accuracy of the obtained coefficients is $|1-T_{N}-R_{N}| \le 1,5 \cdot 10^{-15}$ for each value $N$, that gives the first 14-15 digits as reliable for amplitudes and coefficients)
\label{table.1}}
\end{table}
Here, one can see that the penetrability coefficient $T$ tends to $1,142 \cdot 10^{-18}$ which can be considered as limit value. Parameter $\Delta_{T} = |T_{n+1} - T_{n}|$ demonstrates a convergence of such calculations enough well, which is decreased with increasing of $N$. We include into the table also the amplitudes of transmission and reflection, which have the first some digits stable. We establish (at first time):
\begin{itemize}

\item
The method MIR has a \underline{convergent} algorithm of determination of the amplitudes of wave function, the coefficients of penetrability and reflection which are unique and can be considered as reliable (inside interesting accuracy) concerning to the realistic $\alpha$-nucleus potential in the form BMP.

\item
The method MIR determines the half-live for $^{214}{\rm Po}$ in the interval $\tau_{\rm MIR}=1,55 \cdot 10^{-4} - 1,79 \cdot 10^{-4}$ sec which is essentially closer to experimental value $\tau_{\rm exp}=1,6 \cdot 10^{-4}$ sec in a comparison with half-life in the approach WKB, which equals $\tau_{\rm WKB}=1,1 \cdot 10^{-4}$ sec.
\end{itemize}

\subsection{How is the barrier penetrability changed at increasing of the external boundary $R_{\rm max}$?
\label{sec.5.2}}

Let's analyze whether the convergence in determination of the penetrability coefficient by approach MIR at increasing of the external boundary $R_{\rm max}$ exists. In approach WKB the region of the barrier located between two turning points $R_{2}$ and $R_{3}$ makes up a main part in determination of the penetrability (we obtain: $R_{2}=8,18$~fm and $R_{3}=10,09$~fm). Assuming that approach WKB gives enough well results, we shall suppose that in approach MIR the main region forming the penetrability is the barrier region, while the internal and external parts of the potential do not influence on it practically. Keeping width of each interval to be constant, we shall increase gradually the external boundary $R_{\rm max}$ (by increasing number $N$ of intervals) starting from value near to the external turning point $R_{3}$ ($R_{\rm max} > R_{3}$) and we shall calculate the amplitudes and the penetrability coefficient. These calculated values with different $R_{\rm max}$ are presented in table~\ref{table.2}.
\begin{table}
\begin{center}
\begin{tabular}{|c|c|c|c|c|c|c|c|} \hline
  Boundary &
  \multicolumn{2}{|c|}{Transmission amplitude} &
  \multicolumn{2}{|c|}{Reflection amplitude} &
  \multicolumn{2}{|c|}{Penetrability} &
  Half-live
  \\
  \cline{2-7}
  $R_{\rm max}$, fm &
  $\Re\, (A_{T})$ &
  $\Im\, (A_{T})$ &
  $\Re\, (A_{R})$ &
  $\Im\, (A_{R})$ &
  $T$ &
  $\Delta_{T}$ &
  $\tau_{1/2}$, sec
  \\ \hline
  31 &  1,15E-9  & -3,27E-9  & -0,95509 & 0,29630  & 1,6970E-18 &    -       & 1,17E-4 \\
  40 & -7,94E-10 & -1,47E-9  & -0,95509 & 0,29629  & 1,1258E-18 & 0,5712E-18 & 1,88E-4 \\
  50 & -1,47E-10 & -2,63E-9  & -0,95509 & 0,29629  & 1,1326E-18 & 0,0068E-18 & 1,75E-4 \\
  60 & -9,51E-10 & -1,41E-9  & -0,95509 & 0,29628  & 1,1460E-18 & 0,0066E-18 & 1,73E-4 \\
  70 &  9,56E-10 &  9,87E-10 & -0,95509 & 0,296282 & 1,1463E-18 & 0,0003E-18 & 1,85E-4 \\
  80 & -9,41E-10 & -9,58E-9  & -0,95510 & 0,296282 & 1,1452E-18 & 0,0011E-18 & 1,83E-4 \\
  90 &  4,00E-10 &  1,25E-9  & -0,95510 & 0,296282 & 1,1443E-18 & 0,0009E-18 & 1,70E-4 \\
 100 &  6,36E-10 & -1,13E-9  & -0,95510 & 0,296282 & 1,1428E-18 & 0,0015E-18 & 1,79E-4 \\
  \hline
\end{tabular}
\end{center}
\caption{Dependence of the amplitudes of transmission and reflection, penetrability coefficient and half-live on the external boundary $R_{\rm max}$ for $\alpha$-decay of $^{214}{\rm Po}$
(we use: $R_{\rm min}=0,11$~fm, width of each interval is 0,01 fm (100 intervals inside 1 fm);
$\Delta_{T} = |T_{n+1} - T_{n}|$ is parameter of convergence in calculations of the penetrability coefficient $T$ where $T_{n}$ and $T_{n+1}$ are the penetrability coefficients for each previous and further values of $R_{\rm max}$;
accuracy of the obtained coefficients is $|1-T_{n}-R_{n}| \le 1,5 \cdot 10^{-15}$ for each value of $R_{\rm max}$)
\label{table.2}}
\end{table}
One can see that method MIR gives the convergent value for the penetrability coefficient at increasing of $R_{\rm max}$. But such convergence is not so stable as in increasing of the number of intervals at fixed $R_{\rm max}$. However, starting from $R_{\rm max}=60$~fm, we obtain the first 3 digits stable. Convergence of the amplitude of reflection is higher essentially. We conclude:
\begin{itemize}

\item
The amplitude of transmission and the the penetrability coefficient are convergent with increasing of $R_{\rm max}$ but not so strong, that can be explained by that not only the barrier region but also the external part of the potential up to 60 fm takes place in determination of these characteristics (in contrast with approach WKB).

\item
Half-live obtained by method MIR inside only the barrier region is $\tau_{\rm MIR}=1,17 \cdot 10^{-4}$ sec, which is close to value $\tau_{\rm WKB}=1,1 \cdot 10^{-4}$ sec obtained by approach WKB in~\cite{Buck.1993.ADNDT}. This confirms the conclusion about essential influence of the external part of the potential to the penetrability coefficient determination.
\end{itemize}

\subsection{Dependence of the penetrability on the starting point
\label{sec.5.3}}

\begin{figure}[htbp]
\centerline{\includegraphics[width=54mm]{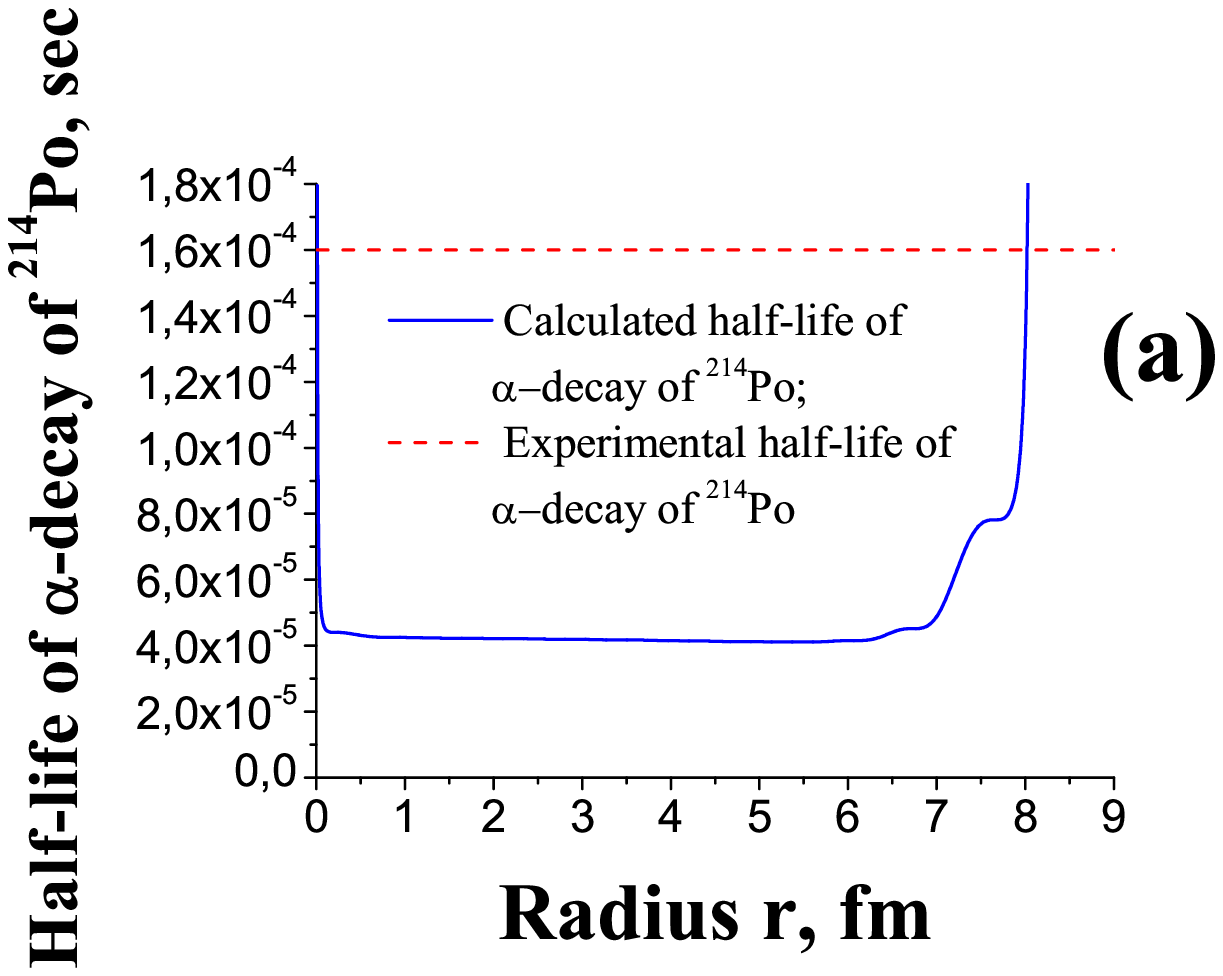}
\includegraphics[width=54mm]{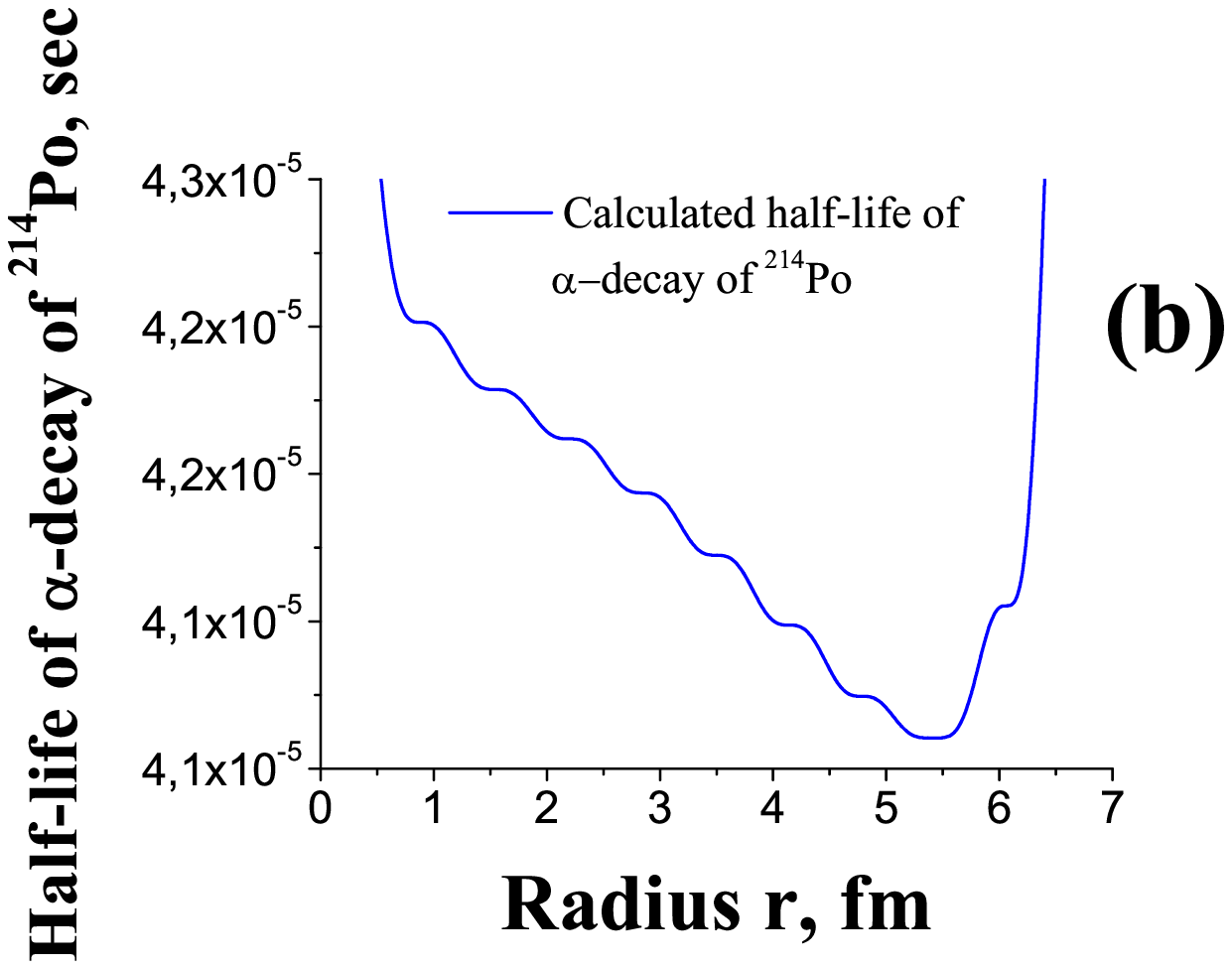}
\includegraphics[width=54mm]{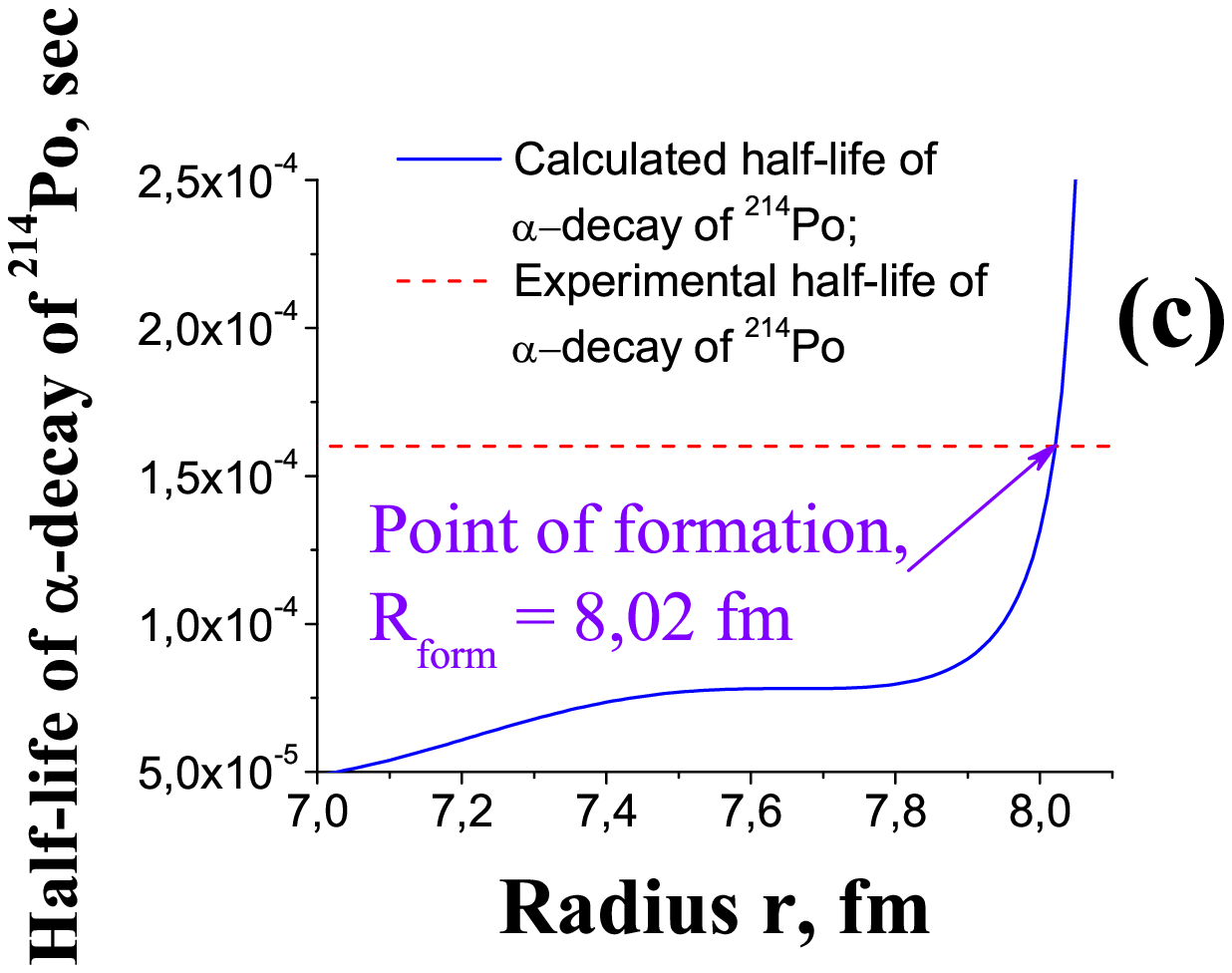}}
\caption{\small
Dependence of half-live for $^{214}{\rm Po}$ on location of the starting point $R_{\rm start}$:
(a) --- dependence of half-live on the starting point from 0 up to 9~fm;
(b) --- increased region where half-live is changed slowly;
(c) --- determination of coordinate of formation of $\alpha$-particle before $\alpha$-decay, as such starting point $R_{\rm start}$, where the calculated half-live is close maximally to its experimental value: we obtain $R_{\rm start} = 8,02$~fm at $\tau_{\rm MIR} = 1,5786 \cdot 10^{-4}$~sec
\label{fig.5.3.1}}
\end{figure}

Now we shall analyze how a location of the internal boundary $R_{\rm min}$ has influence on the penetrability. In the Fig.~\ref{fig.5.3.1} one can see that half-live of $\alpha$-decay of $^{214}{\rm Po}$ is changed in result of displacement of $R_{\rm min}$. Taking into account that width of each interval is 0,01~fm, we shall consider point $R_{\rm min}$ as a \emph{starting point} (with error up to 0,01~fm), from here the $\alpha$-particle begins to move outside (to the right) and is incident to internal side of the barrier starting the first stage of the $\alpha$-decay.
Analyzing figures, we establish
\emph{a (non-constant) dependence of the penetrability coefficient and half-live on the location of the starting point $R_{\rm start}$, where the particle starts to move outside by approach MIR (it has obtained at the first time).}

This dependence opens a new way of determination of the most probable radial coordinate of formation of the $\alpha$-particle before its motion outside in the first stage of the $\alpha$-decay on the basis of a comparable analysis of the calculated half-live with its experimental value: it needs to find such starting point, when the calculated half-live by approach MIR is maximally close to its experimental value. One can assume, that from such space point the $\alpha$-particle begins to move outside in the first stage of the $\alpha$-decay, and there the $\alpha$-particle was formed. We give such definition:

\vspace{2mm}
\noindent
\emph{The radial coordinate $R_{\rm form}$ of the most probable formation of the $\alpha$-particle before $\alpha$-decay is such starting point, at which the calculated half-live $\tau_{MIR}$ is closed maximally to its experimental value $\tau_{\rm exp}$.}

\vspace{2mm}
\noindent
For $\alpha$-decay of $^{214}{\rm Po}$ we obtain $R_{\rm form} = 8,02$~fm (at $R_{2}=8,18$~fm). At such starting point the calculated value of half-live by approach MIR is $\tau_{\rm MIR} = 1,5786 \cdot 10^{-4}$~sec, that is essentially closer to the experimental value $\tau_{\rm exp} = 1,6 \cdot 10^{-4}$~sec for this nucleus, in a comparison with half-live $\tau_{\rm WKB} = 1,1 \cdot 10^{-4}$~sec obtained in~\cite{Buck.1993.ADNDT} by approach WKB.

\begin{figure}[htbp]
\centerline{\includegraphics[width=54mm]{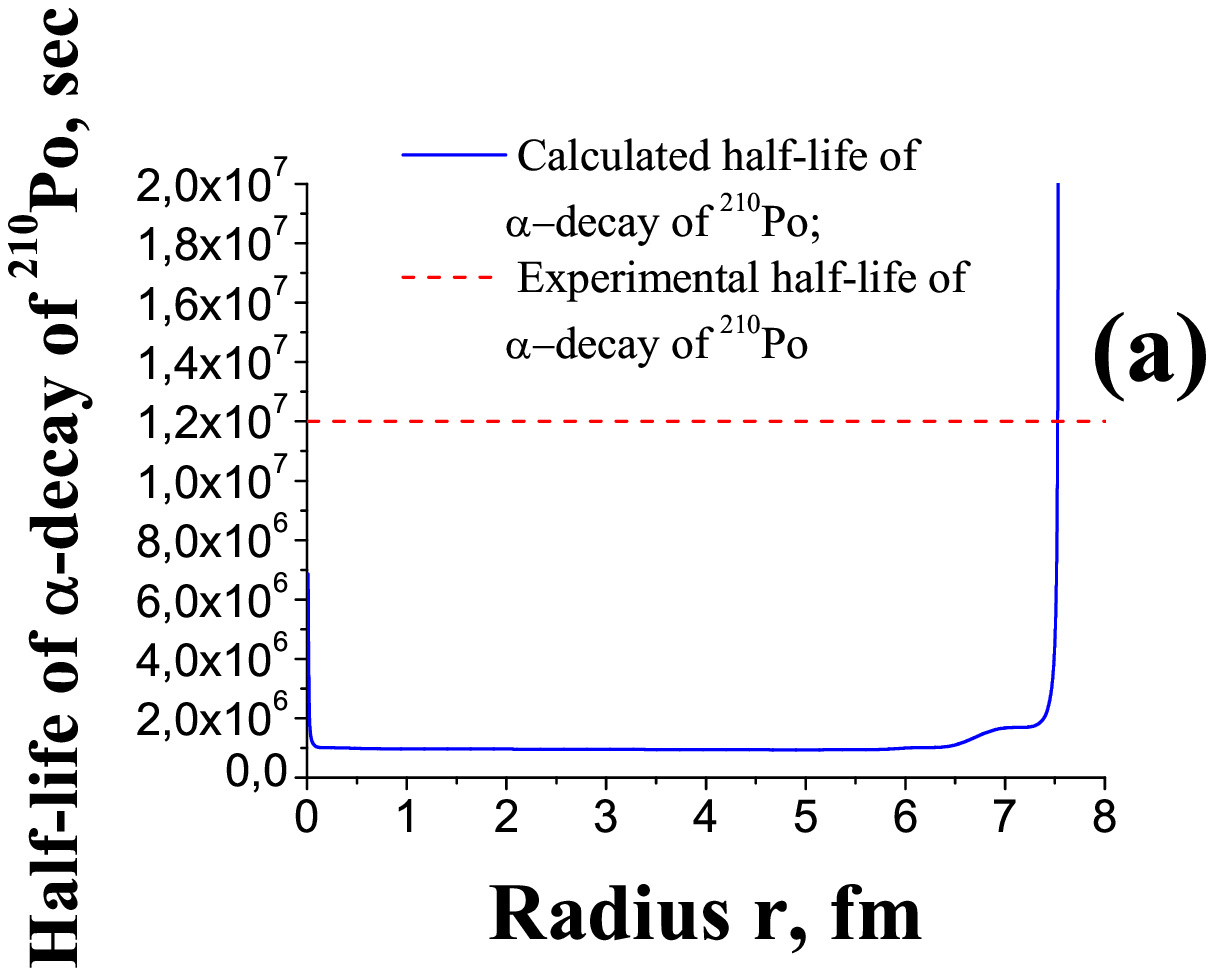}
\includegraphics[width=54mm]{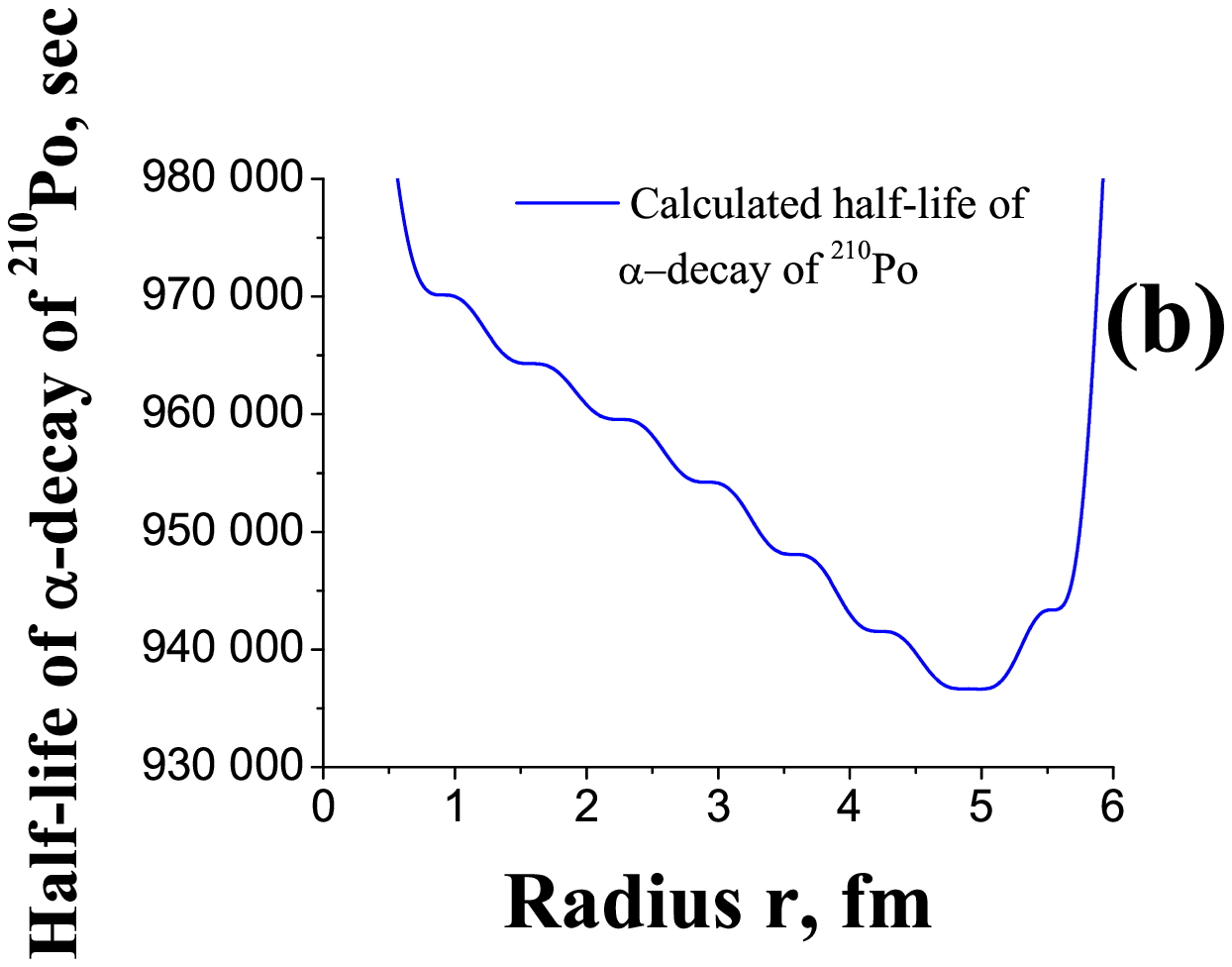}
\includegraphics[width=54mm]{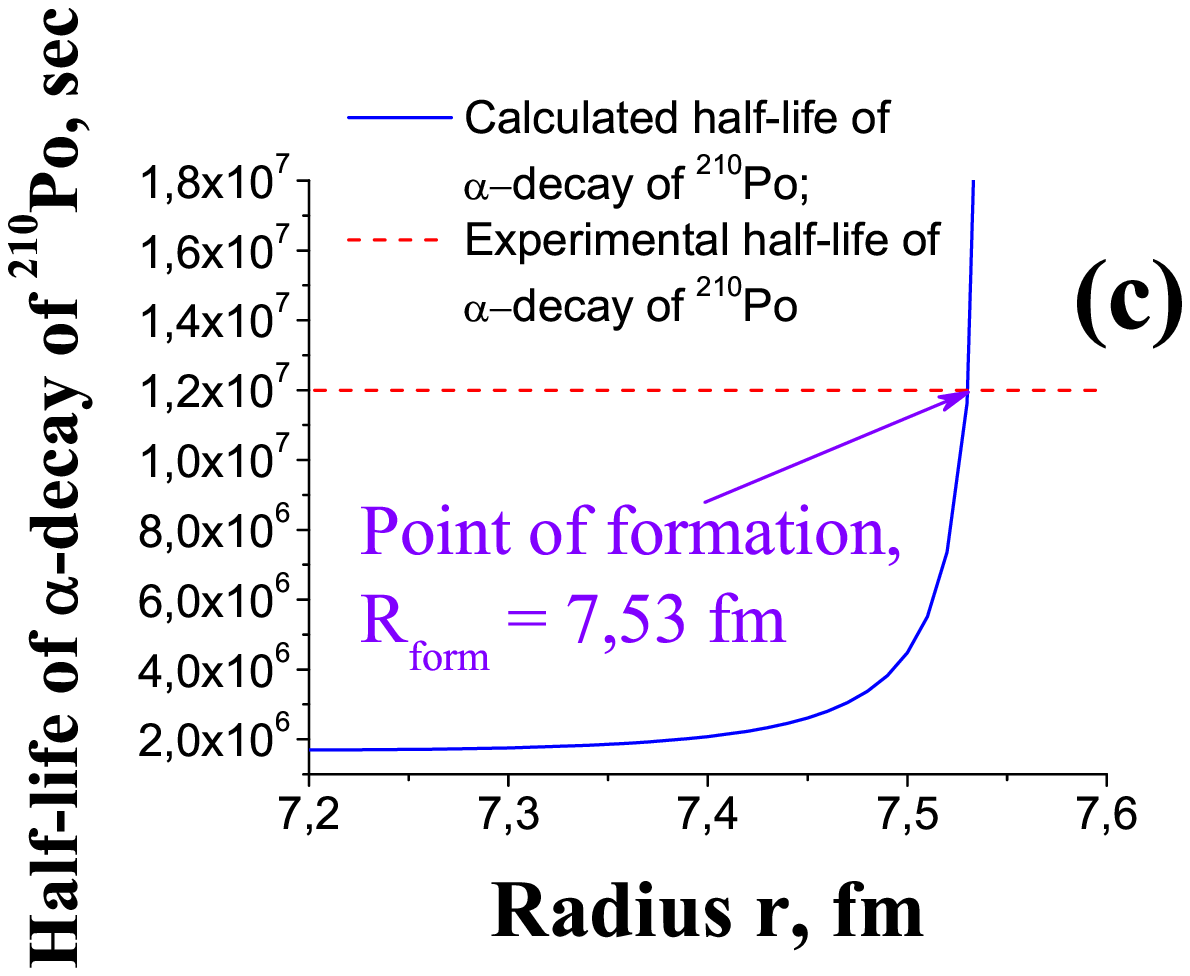}}
\caption{\small
Dependence of half-live for $^{210}{\rm Po}$ on location of the starting point $R_{\rm start}$:
(a) --- dependence of half-live on the starting point from 0 up to 8~fm;
(b) --- increased region where half-live is changed slowly;
(c) --- determination of the coordinate of formation of the $\alpha$-particle before $\alpha$-decay, as such starting point $R_{\rm start}$, where the calculated half-live is close maximally to its experimental value: we obtain $R_{\rm start} = 7,53$~fm at $\tau_{\rm MIR} = 1,16313 \cdot 10^{+7}$~sec
\label{fig.5.3.2}}
\end{figure}

So, on the basis of analysis of $\alpha$-decay of $^{214}{\rm Po}$ we have seen that the \emph{coordinate of formation of the $\alpha$-particle is located inside internal region and is very close to the second turning point $R_{2}$}. But whether is this result accidental, obtained in result of accidental coincidence of parameters of the potential and energy $Q$? Let's analyze another nucleus --- $^{210}{\rm Po}$. While these two nuclei have practically similar shapes of their potentials, their (experimental) half-lives are different essentially. We shall clarify whether our algorithm, used in the method of multiple internal reflections, allows to determine the coordinate of formation of the $\alpha$-particle for $^{210}{\rm Po}$ and where it will be. In Fig.~\ref{fig.5.3.2} one can see that method MIR determines this coordinate simply also and it locates like nucleus $^{214}{\rm Po}$ --- close to the second turning point $R_{2}$.
We obtain value $R_{\rm form} = 7,53$~fm (at $R_{2}=7,642$~fm). At such starting point the calculated value of half-live by approach MIR is $\tau_{\rm MIR} = 1,1631 \cdot 10^{+7}$~sec, that that is essentially closer to the experimental value $\tau_{\rm exp} = 1,2 \cdot 10^{+7}$~sec for this nucleus, in a comparison with half-live $\tau_{\rm WKB} = 3,6 \cdot 10^{+6}$~sec obtained in~\cite{Buck.1993.ADNDT} by approach WKB. So, the found conclusion about formation of the $\alpha$-particle (and also improvement of agreement between the calculated and experimental half-lives after consideration of leaving of the $\alpha$-particle from the coordinate of its formation by approach MIR) is stable both for short-lived and for long-lived nuclei.

%


\subsection{Results of calculation of the coordinate of formation of the $\alpha$-particle for nuclei with charge number $Z=84$
\label{sec.5.4}}

The calculated coordinates $R_{\rm form}$ of formation of the $\alpha$-particle for the $\alpha$-active nuclei with charge number 84 are presented in table~\ref{table.3}.
\begin{table}
\begin{center}
\begin{tabular}{|c|c|c|c|c|c|c|c|c|c|} \hline
 \multicolumn{3}{|c|}{Parent nucleus} &
 \multicolumn{3}{|c|}{Half-live-values, sec} &
 \multicolumn{2}{|c|}{Turning points} &
 \multicolumn{2}{|c|}{$\alpha$-particle formation}
 \\  \hline
  A &
  $Q$, MeV &
  $R_{\rm BMP}$, fm &
  $\tau_{WKB}$ &
  $\tau_{MIR}$ &
  $\tau_{\rm exp}$ &
  $R_{2}$, fm &
  $R_{3}$, fm &
  $R_{\rm form}$, fm &
  $\Delta_{\rm R}$, fm \\ \hline

    201 & 5,830 & 6,956 &            
    6,6\,E4 & 5,7101\,E4 & 5,7\,E4 & 
    7,6350 & 40,4543 &               
    7,5905 & 0,0445 \\               
    202 & 5,733 & 6,959 &            
    1,2\,E5 & 1.3016\,E5 & 1,3\,E5 & 
    7,6363 & 41,1803 &               
    7,5760 & 0,0603 \\               
    203 & 5,524 & 6,965 &            
    2,1\,E6 & 1.9000\,E6 & 1,9\,E6 & 
    7,6385 & 42,7536 &               
    7,5972 & 0,0413 \\               
    204 & 5,516 & 6,965 &            
    1,4\,E6 & 1,9028\,E6 & 1,9\,E6 & 
    7,6384 & 42,8182 &               
    7,5911 & 0,0473 \\               
    205 & 5,356 & 6,970 &            
    1,7\,E7 & 1,0003\,E7 & 1,0\,E7 & 
    7,6405 & 44,0971 &               
    7,5723 & 0,0682 \\               
    206 & 5,358 & 6,970 &            
    9,8\,E6 & 1,4036\,E7 & 1,4\,E7 & 
    7,6406 & 44,0807 &               
    7,5984 & 0,0422 \\               
    207 & 5,248 & 6,973 &            
    6,7\,E7 & 1,0011\,E7 & 1,0\,E8 & 
    7,6418 & 45,0045 &               
    7,6188 & 0,0230 \\               
    208 & 5,248 & 6,973 &            
    4,0\,E7 & 9,0848\,E7 & 9,1\,E7 & 
    7,6415 & 45,0045 &               
    7,6168 & 0,0247 \\               
    209 & 5,007 & 6,980 &            
    1,7\,E9 & 3,1963\,E9 & 3,2\,E9 & 
    7,6444 & 47,1704 &               
    7,6274 & 0,0170 \\               
    210 & 5,439 & 6,966 &            
    3,6\,E6 & 1,2001\,E7 & 1,2\,E7 & 
    7,6379 & 43,4243 &               
    7,6208 & 0,0171 \\               
    212 & 8,985 & 7,380 &               
    1,3\,E-7 & 3,0038\,E-7 & 3,0\,E-7 & 
    8,1658 & 26,2887 &                  
    8,1227 & 0,0431 \\                  
    213 & 8,567 & 7,394 &               
    2,4\,E-6 & 4,1949\,E-6 & 4,2\,E-6 & 
    8,1706 & 27,5711 &                  
    8,1403 & 0,0303 \\                  
    214 & 7,865 & 7,417 &               
    1,1\,E-4 & 1,6014\,E-4 & 1,6\,E-4 & 
    8,1789 & 30,0316 &                  
    8,1053 & 0,0736 \\                  
\hline
\end{tabular}
\end{center}
\caption{Radial coordinates of the most probable formation of the $\alpha$-particle before $\alpha$-decay of nuclei ${\rm Po}$ (with charge number $Z=84$) at different mass numbers $A$
(for each nucleus we use: $R_{\rm min}=0,11$~fm, $R_{\rm max}=100$~fm;
in region from $R_{\rm min}$ to 5~fm number of intervals is 2000,
in region from 5~fm to 7~fm number of intervals is 7000,
in region from 7~fm to $R_{\rm max}$ number of intervals is 1000;
$\Delta_{\rm R} = R_{2} - R_{\rm form}$;
values $Q$, $R_{\rm BMP}$, $\tau_{\rm exp}$ are used from~\cite{Buck.1993.ADNDT}; $\tau_{\rm WKB}$ is half-live determined by approach WKB in~\cite{Buck.1993.ADNDT})
\label{table.3}}
\end{table}
One can see that the coordinate $R_{\rm form}$ is located at small distance $\Delta_{\rm R}$ near the second turning point $R_{2}$. At such choice of the starting point the calculated half-live $\tau_{MIR}$ by approach MIR for each nucleus is essentially closer to the experimental $\tau_{\rm exp}$ in a comparison with half-live $\tau_{WKB}$ obtained in~\cite{Buck.1993.ADNDT} by approach WKB. So, \emph{method MIR is \underline{stable} in determination of the coordinate of formation of the $\alpha$-particle and half-live.} For $\Delta_{\rm R}$ we obtain:
\begin{equation}
  0,01 \, \mbox{\rm fm} \le \Delta_{\rm R} \le 0,07 \, \mbox{\rm fm}.
\label{eq.5.4.1}
\end{equation}

Also we have analyzed possible values of the coordinate of formation of the $\alpha$-particle for these nuclei concerning the potential in the form DI. It turns out that at different locations of the starting point the calculated half-lives by approach WKB are some farther to their experimental values.





\section{Conclusions and perspectives
\label{sec.conclusions}}

In this paper the method of multiple internal reflections in description of the $\alpha$-decay of nucleus in the spherically symmetric approximation is presented. In the approach MIR the formalism of calculation of the amplitudes of wave function, described moving of the $\alpha$-particle from the internal region outside with its tunneling through a realistic radial barrier of arbitrary shape, has been constructed at first time.
We establish (at first time):
\begin{itemize}

\item
The method MIR gives convergent values for the amplitudes of transmission and reflection, the coefficients of penetrability and reflection, obtained in description of leaving of the $\alpha$-particle relatively a potential with barrier in form of a number of rectangular steps (multi-steps potential), with tending of this potential to the realistic $\alpha$-nucleus potential. Therefore, limit values of the amplitudes and coefficients can be considered as the exact values concerning the realistic $\alpha$-nucleus potential (without application of WKB approximation).

\item
Error in determination of the coefficients and amplitudes by the method MIR concerning the realistic $\alpha$-nucleus potential is determined by accuracy of description of this potential by the multi-steps potential and accuracy of computer calculations (in all calculations, we have $|T+R-1| < 1,5 \cdot 10^{-15}$).

\end{itemize}
In approach MIR a dependence of the penetrability coefficient and half-live on the location of the starting point $R_{\rm start}$, from where $\alpha$-particle begins to move outside, is observed. According to this property:
\begin{itemize}

\item
We define \emph{the radial coordinate of the most probable formation of the $\alpha$-particle} inside nucleus before its $\alpha$-decay as such starting point, at moving of the $\alpha$-particle from which outside the half-live, calculated by approach MIR, is maximally closed to its experimental value.

\item
For a number of nuclei with the same charge number $Z=84$ we have obtained essentially closer values for half-live, calculated by such approach, to their experimental values in a comparison with half-lives obtained in \cite{Buck.1993.ADNDT} by WKB approach. This result is stable for all studied nuclei.

\item
Inaccuracy of determination of half-lives for arbitrary studied nucleus on the basis of WKB method in a comparison with its experimental value can be explained (first of all) by reduction of the above-barrier external part of the radial potential starting from the external turning point, while inclusion of such contribution (by the method MIR, up to 50-60~fm) improves essentially the agreement with the experimental half-live.

\end{itemize}
One can consider the pointed above dependence of half-live on the starting point by approach MIR as a new method of determination of the radial coordinate of the most probable formation of the $\alpha$-particle before the $\alpha$-decay of the nucleus on the basis of analysis of the experimental half-live for this nucleus.

After construction of the algorithm of exact determination of the amplitudes of wave function relatively the multi-steps barrier (where these steps have arbitrary shapes), the method of multiple internal reflection transforms into a power tool for determination of stationary and (non-stationary) characteristics with high accuracy (and without application of WKB approximation) in variety of spherically symmetric problems of scattering and decay. Being the method with exact solutions, it allows to study deeper unusual and amazing properties of quantum systems and potentials, that points out availability of further improvements of the method, and such researches can have physical sense.
So, it can be interesting by use of this method to study a \emph{reflectionless property}, which is most evidentally shown in simplest soliton-like potentials \cite{Cooper.1995.PRPLC} (see also \cite{Zakhariev.1994.PEPAN,Zakhariev.1999.PEPAN,Zakhariev.2002.PEPAN}) and shape invariant potentials \cite{Shabat.1992.INPEE,Spiridonov.1992.PRLTA},
to clarify whether the \emph{effect of reflectionless tunneling} is possible in real (hermitian) radial potentials \cite{Maydanyuk.2005.APNYA,Maydanyuk.2005.Surveys_in_HEP}.
We suppose that this method should open a way for construction of new types of space asymmetric deformations of 1D reflectionless potentials \cite{Maydanyuk.arXiv:0710.4062},
will give new variants of angular deformations of 3D reflectionless potentials obtained in frameworks of group theory \cite{Kerimov.2006.JMP,Kerimov.2007.JPA}.
A generalization of the method MIR into complex (non-hermitian) potentials should allow to study properties of barriers in scattering problems with these potentials studied by approaches of non-linear supersymmetry (for example, see~\cite{Andrianov.2007.NPA,Sokolov.2007.NPA,Andrianov.2003.NuclPhys,Andrianov.2004.JPAGB,Baye.1996.NPA})
and N-fold supersymmetry \cite{Aoyama.2001.NPB,Aoyama.2001.PLB,Tanaka.2002.JMP,Aoyama.2006.AP,Tanaka.2006.JPA},
by approaches of hidden supersymmetry \cite{Plyushchay.2007.AP.v322,Plyushchay.2007.JPA}.
In application to such problems, the method MIR can be a good test of early obtained results on the basis of methods of SUSY QM, group theory and inverse problem approach.

\bibliography{Alpha_decay_MIR_eng}

\end{document}